\documentclass[]{mn2e}

\usepackage{amssymb}
\usepackage{amsmath}
\usepackage{graphicx}
\usepackage{appendix}
\usepackage{longtable}
\usepackage{url}

\title[Comparing PyMorph and SDSS photometry. II.]
      {Comparing PyMorph and SDSS photometry. II.\\
       The differences are more than semantics and are not dominated by intracluster light}
\author[Bernardi et al.]{\parbox{\textwidth}{M. Bernardi$^{1}$\thanks{E-mail: bernardm@sas.upenn.edu}, J.-L. Fischer$^{1}$, R. K. Sheth$^{1}$, A. Meert$^{1}$, M. Huertas-Company$^{2}$, \\ F. Shankar$^{3}$ \& V. Vikram$^{1}$} \vspace{0.4cm}\\
\parbox{\textwidth}{$^{1}$Department of Physics and Astronomy, 
                    University of Pennsylvania, Philadelphia, PA 19104, USA\\
$^{2}$Observatoire de Paris (GEPI), CNRS and Universite' Paris Diderot, 
                    4 Rue Thomas Mann, 75013 Paris, France\\
$^{3}$School of Physics and Astronomy, University of Southampton,
                    Southampton SO17 1BJ, UK\\                    
}}

\begin{document}
 \date{Accepted .  Received ; in original form }

\maketitle

\label{firstpage}

\begin{abstract}
  The Sloan Digital Sky Survey pipeline photometry underestimates the brightnesses of the most luminous galaxies. This is mainly because (i) the SDSS overestimates the sky background and (ii) single or two-component Sersic-based models better fit the surface brightness profile of galaxies, especially at high luminosities, than does the de Vaucouleurs model used by the SDSS pipeline.  We use the {\tt PyMorph} photometric reductions to isolate effect (ii) and show that it is the same in the full sample as in small group environments, and for satellites in the most massive clusters as well.  None of these are expected to be significantly affected by intracluster light (ICL).  We only see an additional effect for centrals in the most massive halos, but we argue that even this is not dominated by ICL. Hence, for the vast majority of galaxies, the differences between {\tt PyMorph} and SDSS pipeline photometry cannot be ascribed to the semantics of whether or not one includes the ICL when describing the stellar mass of massive galaxies. Rather, they  likely reflect differences in star formation or assembly histories.
  Failure to account for the SDSS underestimate has significantly biased most previous estimates of the SDSS luminosity and stellar mass functions, and therefore Halo Model estimates of the $z\sim 0.1$ relation between the mass of a halo and that of the galaxy at its center.  
 We also show that when one studies correlations, at fixed group mass, with a quantity which was not used to define the groups, then selection effects appear.
We show why such effects arise, and should not be mistaken for physical effects.
\end{abstract}


\section{Introduction}
The observed magnitudes reported by the Sloan Digital Sky Survey are underestimated especially at the highest luminosities (e.g. Bernardi et al. 2007; von der Linden et al. 2007; Bernardi et al. 2013; Meert et al. 2015; D'Souza et al. 2015; Bernardi et al. 2017).  In a companion paper \cite{F16}, we show that this is due to a combination of sky background and model fitting effects: (i) the SDSS overestimates the sky background (Blanton et al. 2005; Bernardi et al. 2007; Hyde \& Bernardi 2009; Blanton et al. 2011; Meert et al. 2015), and (ii) single- or two-component Sersic model based estimates of galaxy luminosities are more reliable than estimates based on single exponential, single de Vaucouleurs, or a linear combination of the best fitting exponential and de Vaucouleurs models (the so-called {\tt cModel} magnitudes) used by the SDSS pipeline (e.g. Bernardi et al. 2010; Bernardi et al. 2014).  Fischer et al. show that PyMorph sky estimates are fainter than those of the SDSS DR7 or DR9 pipelines, but are in excellent agreement with the estimates of Blanton et al. (2011).  The PyMorph Sersic-based estimates \cite{m13,m15,m16} are more reliable and return more light than do SDSS estimates, and the difference is most pronounced at the highest luminosities.  This can have a dramatic impact on the estimated $z\sim 0.1$ luminosity and stellar mass functions \cite{b13,b16,b17}, and hence on Halo Model \cite{cs02} based estimates of the relation between stellar and halo mass (Kravtsov et al. 2014; Shankar et al. 2014).  They also impact models for the formation of the massive galaxies because they affect estimates of the mass scale on which galaxy scaling relations show curvature \cite{b11}.

The most luminous galaxies reside in or at the centers of clusters.  Since clusters are known to possess intercluster light (hereafter ICL), it is natural to ask if the PyMorph Sersic-based photometry, with its improved sky estimates, is brighter than the SDSS measurements (i.e. {\tt Model} or {\tt cModel} magnitudes) because it includes more of this ICL component.  Previous work, based on stacked images of central galaxies of massive clusters, suggests that the ICL only contributes substantially to the surface brightness profile on scales larger than 50~kpc \cite{zibettiICL}, where the surface brightness has dropped below about 27~mags~arcsec$^{-2}$.  A stacking analysis of LRGs suggests this scale may be even larger for lower mass groups \cite{lrgICL}.  For the vast majority of central galaxies in the SDSS, this corresponds to a scale where the surface brightness has dropped to less than 1\% of the sky brightness, making it extremely difficult to detect in individual images.  While this means that it is very unlikely that the ICL contributes substantially to the PyMorph reductions, in what follows, we provide additional evidence that the PyMorph-SDSS difference is unlikely to be due to ICL.  

Our logic is simple:  Since the ICL should be centered on the cluster center, and is expected to be fainter at larger cluster-centric distances, it is reasonable to suppose that it will affect the photometry of the central galaxy more than the satellites. The ICL is also expected to be fainter around central galaxies in less massive groups. Therefore, we study if the difference between PyMorph and SDSS magnitudes depends on whether the galaxy is a central or a satellite, as well as if it depends on cluster mass.  Section~2 describes our galaxy sample and
 shows why the ICL is unlikely to play a major role in the images of individual galaxies.  Section~3 describes the group catalogs we use for separating centrals from satellites.  Section~4 presents the main results of our analysis and Section~5 summarizes our findings. While the main text presents the results using the Sersic-Exponential photometry, Appendix~\ref{sersic} shows a similar analysis for the single-component Sersic fits since these are much more common in the literature. Finally, Appendix~\ref{secbias} discusses a pernicious selection effect which arises when working with group catalogs, and which we were careful to avoid. 

When necessary, we assume a spatially flat background cosmology with parameters $(\Omega_m,\Omega_\Lambda)=(0.3,0.7)$, and a Hubble constant at the present time of $H_0=70$~km~s$^{-1}$Mpc$^{-1}$.

\section{Comparison of PyMorph and SDSS luminosities:  The main galaxy sample}

\subsection{The parent galaxy sample}
The analysis which follows is based on the Main Galaxy sample of the Ninth Data Release of the Sloan Digital Sky Survey (hereafter SDSS DR9; Aihara et al. 2011); the sample is limited to a Petrosian $r$-band apparent magnitude of $m_{r{\rm Pet}}\le 17.7$.  (Our conclusions are unchanged if we use DR7 values instead of DR9.)  To about $\sim 660,000$ of these (those at $z\le 0.25$),  Huertas Company et al. (2011) have assigned Bayesian Automated Morphological Classification weights which represent the probabilities that the galaxy is Elliptical, S0, Sab, or Scd.  The vast majority of the most luminous galaxies are Es or S0s, so, in what follows, we define the E+S0 weight to equal $p$(E+S0) $\equiv p$(E) + $p$(S0).

For every DR9 galaxy, the SDSS pipeline provides {\tt Model} magnitudes, which are based on separately fitting an exponential and a de Vaucouleurs form to the surface brightness profile and choosing the value returned by the model which fits best.  Hence, for essentially all E+S0s, the {\tt Model} magnitude is that from the de Vaucouleurs fit. (Our conclusions are unchanged if we use {\tt cModel} magnitudes).

We will compare these {\tt Model} magnitudes with single-component Sersic and two-component Sersic-Exponential fits (hereafter {\tt Ser} and {\tt SerExp}) to these same objects provided by Meert et al. (2015).  These {\tt Ser} and {\tt SerExp} fits are returned by the PyMorph algorithm which is described and tested in Meert et al. (2013, 2015, 2016) and used by Bernardi et al. (2013, 2014, 2016, 2017).  Although those tests were based on DR7 objects, Fischer et al. (2017) show that the difference between DR7 and DR9 is negligible for PyMorph.  I.e., the parameters provided by Meert et al. (2015) for DR7 can also be used for DR9.  The two-component {\tt SerExp} fits are the most accurate (i.e. least biased) of the PyMorph outputs \cite{m13,b14}.  For this reason, we use these in the main text.  However, single-component fits are much more common in the literature; we show our analysis of {\tt Ser} photometry in Appendix~\ref{sersic}.

We also present our results as a function of stellar mass $M_*$.  For a given luminosity (e.g., {\tt Model}, or {\tt SerExp}), we obtain $M_*$ by multiplying $L$ by the dust-free $M_*/L$ estimates of Mendel et al. (2014) assuming a Chabrier (2003) IMF.  Bernardi et al. (2017) describe a number of other reasonable choices of $M_*/L$.  The results which follow are robust to changes in this choice.  

\subsection{Example profiles and fits}\label{images}
Before showing results of a statistical analysis, it is useful to look at a few representative images.  This helps build intuition for what it is that we will be averaging.  Our aim here is to directly address what previous work based on stacked images imply for our analysis of individual images.  We are particularly interested in the effects of the ICL.  If ICL is defined as being an excess above the best-fitting {\tt deV} profile, then it is only expected to contribute significantly on scales larger than 50~kpc \cite{zibettiICL}.  On the other hand, if ICL is a departure from the best-fitting {\tt Ser} profile, then the relevant scale may be about a factor of two larger \cite{lrgICL}.

\begin{figure*}
  \centering
  \includegraphics[width=0.95\textwidth]{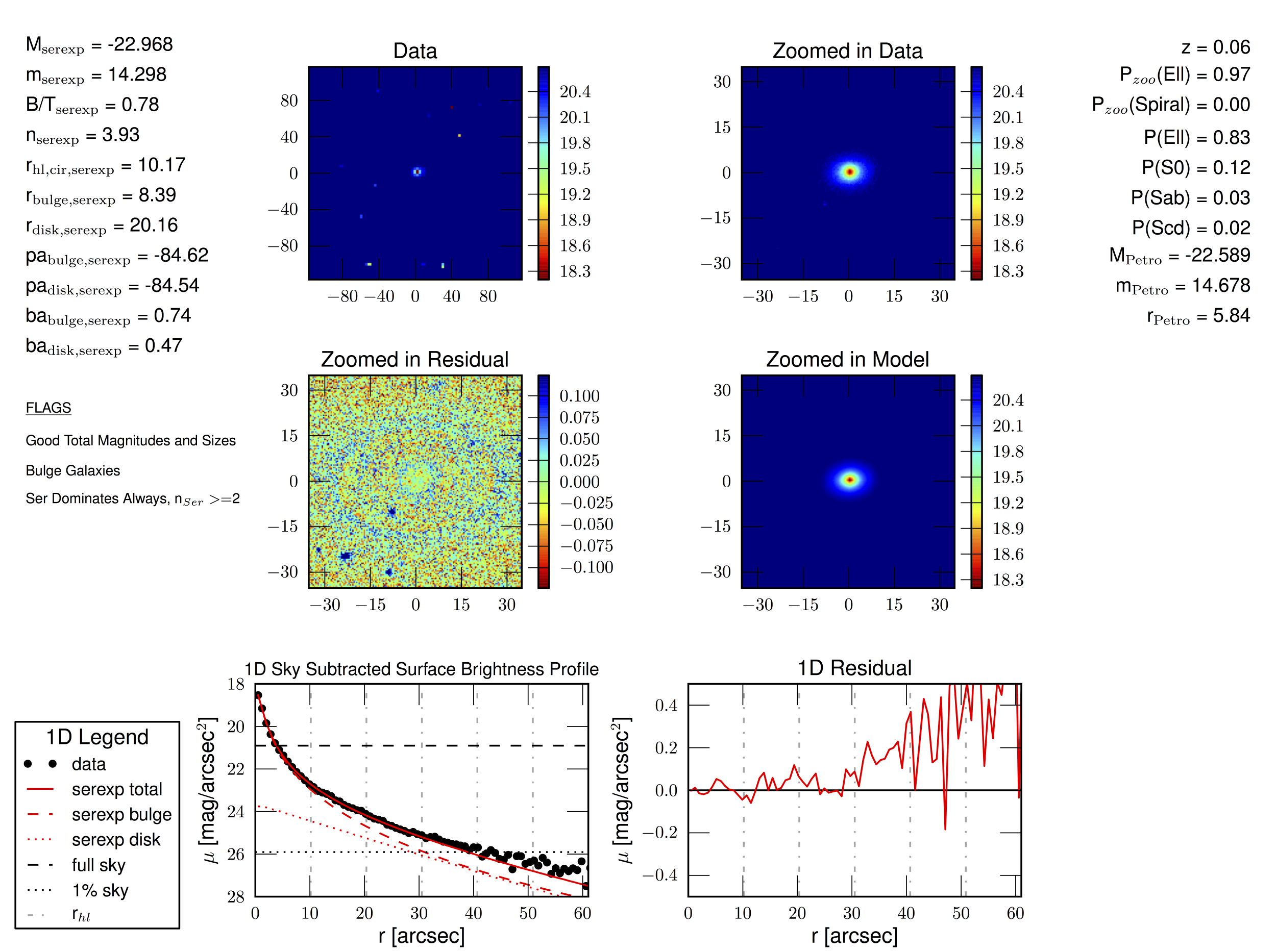}
  \caption{Image (top panels) and 2D {\tt SerExp} fit and residuals (middle panels) of an object for which the {\tt SerExp} fit indicates has a half light radius $R_{hl} = 12.68$~kpc. The axis labels are in arcsecs and the color scale is in mags/arcsec$^2$ (the residuals are computed as fit-data). The legends on either side provide a wealth of information about the parameters of the fit. The 2D data images are shown with background sky included, while the 1D angular average profile shown in the bottom panel (black points) is computed using the background subtracted data. The dot-dashed vertical lines in the bottom panels show the half light radius ($r_{hl}$ in arcsec) and its multiples. Dashed and dotted horizontal lines show the measured sky level and 1\% of its value, respectively. The surface brightness profile of this galaxy drops to 1\% of sky at a scale which is about $4\times R_{hl}$. Whereas the fitting is done in 2D, and accounts for the profiles of the other objects in the field, the residuals -- and the 1D angular averages shown in the bottom panel -- do not.  Hence, one should resist the temptation to associate the fact that the data in the bottom panels are slightly brighter than the {\tt SerExp} fit with ICL; some of the apparent excess is due to the extended profiles of the other objects in the group, rather than ICL.  A single {\tt Ser} fit to this object has $n_{\rm Ser} = 5$. The bulge component of the {\tt SerExp} fit is very similar to a single {\tt deV} fit. The departure from a {\tt deV} profile is observed at $\sim 1R_{hl} \sim 13$~kpc which is much smaller than the 50-100 kpc scale expected for the ICL.}
 \label{prof1}
\end{figure*}

\begin{figure*}
  \centering
  \includegraphics[width=0.95\textwidth]{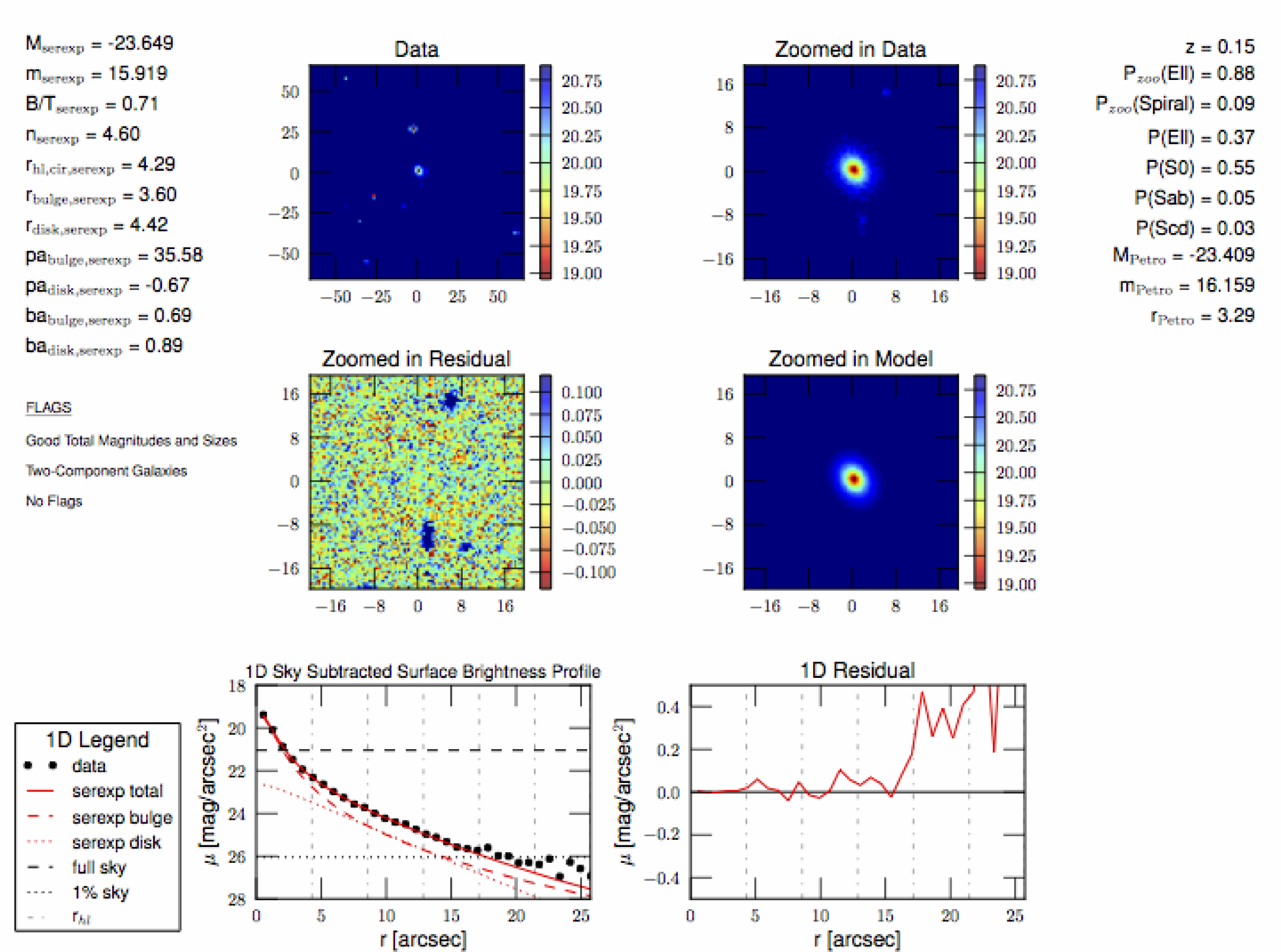}
  \caption{Same as previous figure, but now for a more luminous object which is more distant. In this case, the {\tt SerExp} fit has $R_{hl} = 13.37$~kpc; $R_{1\%{\rm sky}}$ is $4\times$ larger.  Inside $R_{1\%{sky}}$, a single {\tt Ser} profile with $n_{\rm Ser} = 5.2$ also provides a good fit (not shown).}
 \label{prof2}
\end{figure*}

\begin{figure*}
  \centering
  \includegraphics[width=0.95\textwidth]{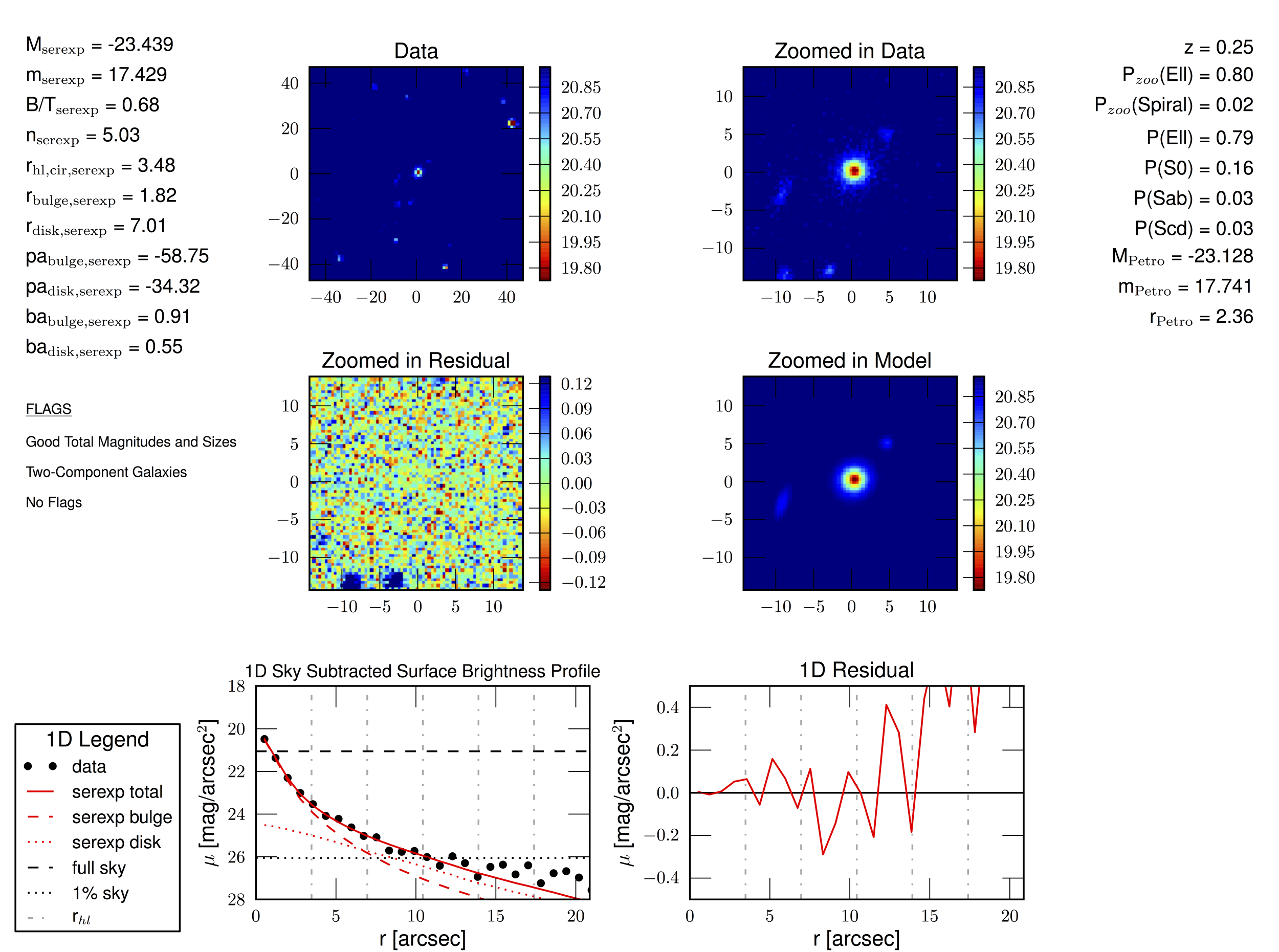}
  \caption{Same as previous figure, but now for an object of similar luminosity which is even more distant. In this case, $R_{hl} = 18.07$~kpc and $R_{1\%{\rm sky}}$ is $3\times$ larger. A single {\tt Ser} profile with $n_{\rm Ser} = 7.9$ provides a good fit inside $R_{1\%{sky}}$ (not shown).}
 \label{prof3}
\end{figure*}

Figures~\ref{prof1}--\ref{prof3} show the surface brightness profiles of three luminous ($M_r \le -23$) galaxies in the SDSS main galaxy sample, which are at a range of distances ($z = 0.06$, 0.15 and 0.25).  As our primary interest here is in the role of the ICL, all three objects are central galaxies in the group catalogs we use extensively later in this paper.  In all cases, the top panels show the image, the middle panels show the best 2D {\tt PyMorph SerExp} fit and residuals from it, and the bottom panels show 1D angular averages of the profile, the fit and residuals (the latter are computed as fit-data).  The 2D residuals show the image including sky whereas those in the bottom the sky has been subtracted.  These bottom panels are for illustration only, as the fitting was done in 2D with more care taken to remove light associated with the other objects in the field.  The legends on either side provide a wealth of information about the parameters of the fit (see Meert et al. 2015 for details).  In all cases, the image cutouts are labeled in units of arcsecs, and the color scale is in mags/arcsec$^2$.

Figure~\ref{prof1} shows a galaxy which was selected because of its large apparent brightness ($m_r \sim 14.3$), so that the dynamic range between the half light radius $R_{hl}$ and the scale $R_{1\%{\rm sky}}$ on which the profile has dropped to 1\% of sky, is large.  The {\tt SerExp} absolute magnitude of this object is $\sim -23$ and $R_{hl} \approx 13$kpc.
For this object, $R_{1\%{\rm sky}}\approx 50$~kpc; note that there is no obvious feature in the profile shape on scales smaller than this.  Indeed, a single {\tt Ser} fit to this object has $n_{\rm Ser}=5$ and is not very different from the best fitting {\tt SerExp} profile shown by the solid curve in the bottom left panel.  The dashed curve in the same panel shows the bulge component of the {\tt SerExp} fit (a single Sersic profile with $n\approx 4$, meaning that this bulge component has a {\tt deV} profile).  Beyond about 10~arcsec from the center -- i.e., beyond about $R_{hl}$ -- the second component (dotted curve) is clearly necessary.  Note in particular that this second component, which describes light in excess of a {\tt deV} profile, is necessary on scales which are much smaller than the 50~kpc associated with the ICL.

Figure~\ref{prof2} shows a more luminous ($M_r=-23.65$) and distant ($z=0.15$) central galaxy with $m_r \approx 16$.  For this object too, the bulge component is not too different from a {\tt deV} profile, but the second component is necessary even on scales as small as 15~kpc.  A single {\tt Ser} profile with $n=5.2$ also provides a good fit on all scales smaller than $R_{1\%{\rm sky}}$.

\begin{figure}
  \centering
  \includegraphics[scale = .42]{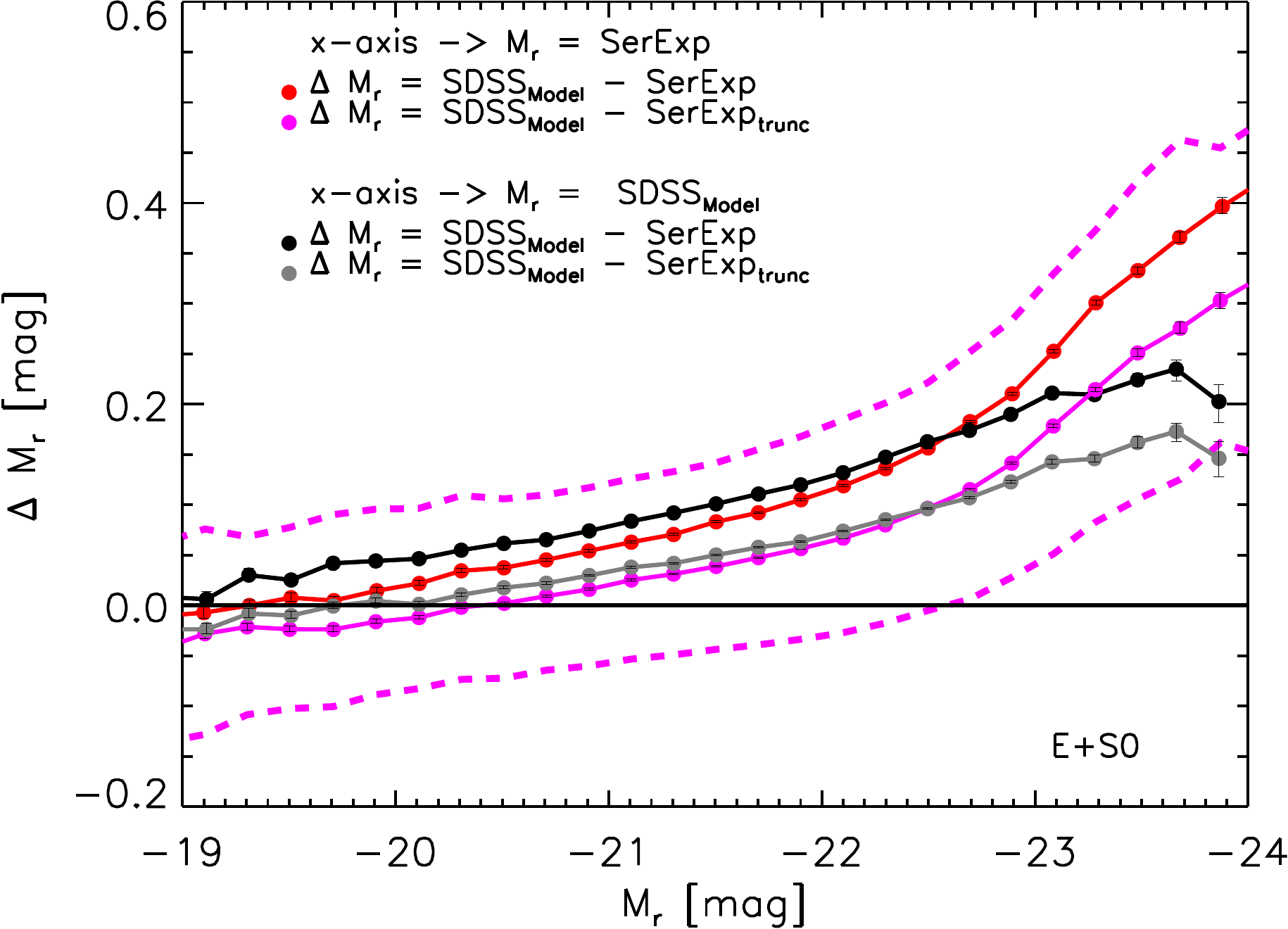}
  \caption{At high luminosites, the mean magnitude difference between SDSS {\tt Model} and PyMorph {\tt SerExp} magnitudes depends strongly on which quantity is used as reference.  Truncating the PyMorph magnitudes similarly to what is done for the SDSS makes them fainter by $\le 0.1$~mags on average; it is not the dominant effect.}
 \label{whichMag}
\end{figure}

Figure~\ref{prof3} shows another object of similar luminosity, but at even higher redshift, so its apparent magnitude is significantly fainter ($m_r \sim 17.4$).  Comparison with the bottom left panels of the previous figures shows the reduced dynamic range which makes it more difficult to detect departures from a single {\tt Ser} fit which, in this case, has $n_{\rm Ser} = 7.9$.  As for the previous two examples, the bulge is closer to {\tt deV}, and the need for a second component is already evident on scales of order 20~kpc.  

Thus, to the extent that these galaxies are representative of all central galaxies, these figures make two points.  First, departures from a {\tt deV} profile are almost always detected with high significance (see also e.g. Gonzalez et al. 2005; Bernardi et al. 2007).  Moreover, a pure {\tt deV} profile becomes a poor fit on much smaller scales than is reasonable to associate with the ICL.  Second, if the ICL is associated with departures from a single {\tt Ser} (rather than {\tt deV}) fit, then if this occurs, it is at surface brightnesses which are too faint to be seen in individual images.  Therefore, the departures from a pure {\tt deV} profile, which {\tt PyMorph} detects in its {\tt Ser} or {\tt SerExp} fits and which are the subject of this paper, likely reflect differences in star formation or assembly histories, rather than ICL.

Having illustrated that differences from {\tt deV} photometry are common, we now turn to a statistical analysis of these differences.  

\subsection{Comparison of SDSS and PyMorph pipelines}

Figure~\ref{whichMag} shows a comparison of SDSS {\tt Model} and PyMorph {\tt SerExp} magnitudes; a similar analysis using {\tt Ser} magnitudes is shown in Appendix~\ref{sersic}.  Here, as in most of the figures which follow, objects have been weighted by $p$(E+S0).  This ensures that we are working with a sample for which {\tt Model} = de Vaucouleurs and removes the question of how the morphological mix affects the PyMorph-SDSS comparison.  

Symbols with error bars show the median and the error on it (only bins with more than 50 objects are shown), and dashed curves show the region which encloses 68\% of the sample in each absolute magnitude bin.  The SDSS {\tt Model} magnitudes are increasingly fainter as luminosity increases.  Some of the differences in Figure~\ref{whichMag} are simply due to the fact that SDSS magnitudes are based on integrating the fitted profile to approximately $7.5\times$ the semi-major axis $a_e$, whereas PyMorph does not truncate.  For this reason, we have shown the result of truncating the PyMorph fits as well:  this makes them fainter (by less than 0.1~mags on average), but other than this shift, the overall trends with luminosity are unchanged.  Therefore, truncation is not the primary reason why SDSS is fainter (see Sections 2.3 and 2.4 in Fischer et al. 2017 for more discussion of truncation).  

Fischer et al. (2017) show that, once truncation has been accounted for, there are two effects which contribute to biasing SDSS low (rather than biasing PyMorph high).  These are due to differences in how the background sky is estimated, and what model is fit to the surface brightness profile. Regarding the first effect, there is now general consensus that the SDSS treatment of the sky is flawed. This affects nearby galaxies (as emphasized by Blanton et al. 2011) but also high luminosity galaxies which have relatively large angular sizes and tend to be in crowded fields (Fischer et al. 2017 and references therein). PyMorph attributes less light to the sky than does the SDSS; as a result, it assigns more light to the galaxy than does the SDSS. Fischer et al. (2017) also show that PyMorph sky estimates are in excellent agreement with those of Blanton et al. (2011). About half of the bias at high luminosities arises from fitting different models (the second effect). At the high luminosity end, most galaxies are E+S0s, for which the SDSS {\tt Model} magnitudes are essentially {\tt deV} magnitudes, and {\tt deV} fits return less light than {\tt SerExp} or {\tt Ser} fits.

\section{Galaxy groups in the SDSS}
Our goal is to check if the differences between PyMorph and {\tt Model} photometry shown in Figure~\ref{whichMag} depend on whether a galaxy is a central or a satellite.  To achieve our goal, we use two group catalogs in which centrals and satellites have been identified.  One of these is from Yang et al. (2007) (the DR7 version, hereafter {\tt Yang+}), and the other is the {\tt redMaPPer} sample of Rykoff et al. (2014).  Whereas the former identifies groups spanning a wide range of masses, the latter only identifies the very most massive clusters (the {\tt redMapper} algorithm is not well-suited for identifying lower mass/richness groups).

\subsection{Description of group catalogs}

The {\tt redMapper} sample is drawn from the SDSS DR8 footprint which covers $\sim 10,000$ deg$^2$.  Groups are identified on the basis of angular positions and color.  Of the $1.7\times 10^6$ objects in $2.6\times 10^4$ groups in the {\tt redMaPPer} sample, Meert et al. only provide PyMorph reductions for $1.8\times 10^4$:  these are the subset which have spectroscopic information and were in the SDSS DR7 $\sim 7,700$ deg$^2$ footprint.  These objects are in about 3,400 clusters, of which about 2,400 are at $z\le 0.25$, and 1,200 are at $z\le 0.2$.  

In contrast, {\tt Yang+} identify $6 \times 10^5$ galaxies in about $4.7 \times 10^5$ groups at $z\le 0.2$ in the SDSS DR7 footprint.  Most of these are much less massive groups, of course.  {\tt Yang+} defined their catalog using a complex iterative procedure that makes use of observed angular positions, redshifts and {\tt Model} photometry.  They estimate the halo masses $M_{\rm Halo}$ of their groups using two simple proxies for halo mass.  One is based on summing up the {\tt Model} luminosities of the central+satellite galaxies.  As a result, there is a tight correlation between $M_{\rm Halo}$ and {\tt Model} which will be important below (see also Appendix~\ref{secbias}).  The other is based on summing their stellar masses, which they estimate by multiplying the {\tt Model} luminosities by a simple color-based estimate of $M_*/L$ (see their equation~2). We use their luminosity based estimate, but the results which follow are robust to changes in this choice. 

\begin{figure}
 \centering
 \includegraphics[scale = .42]{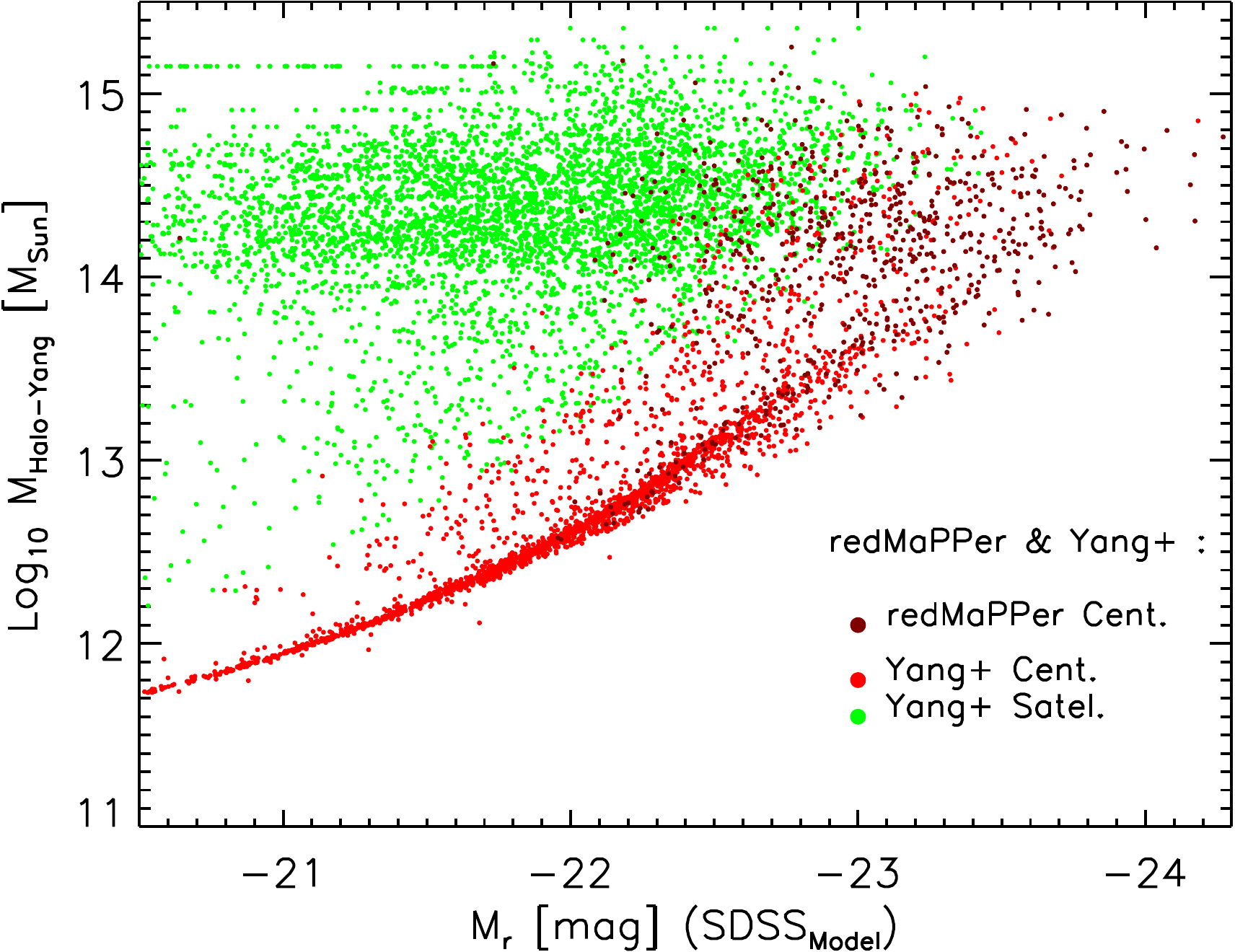}
 \caption{Distribution of E+S0s which are in both the {\tt Yang+} and the {\tt redMaPPer} catalogs.  The {\tt Yang+} definition of $M_{\rm Halo}$ means that there is a tight correlation between $M_{\rm Halo}$ and {\tt Model} magnitudes especially at smaller masses. Many of the objects which {\tt Yang+} classify as being centrals in groups less massive than $10^{14}M_\odot$, are called satellites by {\tt redMaPPer} (the {\tt redMaPPer} satellites are all the points which are not brown). Only above $10^{14}M_\odot$ do the two groups agree on the central-satellite classification.}
 \label{yangRedmapper}
\end{figure}

The {\tt Yang+} catalog only extends out to $z=0.2$, so our first step was to identify objects which appear in the {\tt redMaPPer} catalog as well.  There are 13,253 such objects of which  8,081 have $p$(E+S0)$\ge 0.7$.  Figure~\ref{yangRedmapper} shows the distribution of these objects in the $M_{\rm Halo}-${\tt Model} magnitude plane.

\begin{figure}
  \centering
  \includegraphics[scale = .42]{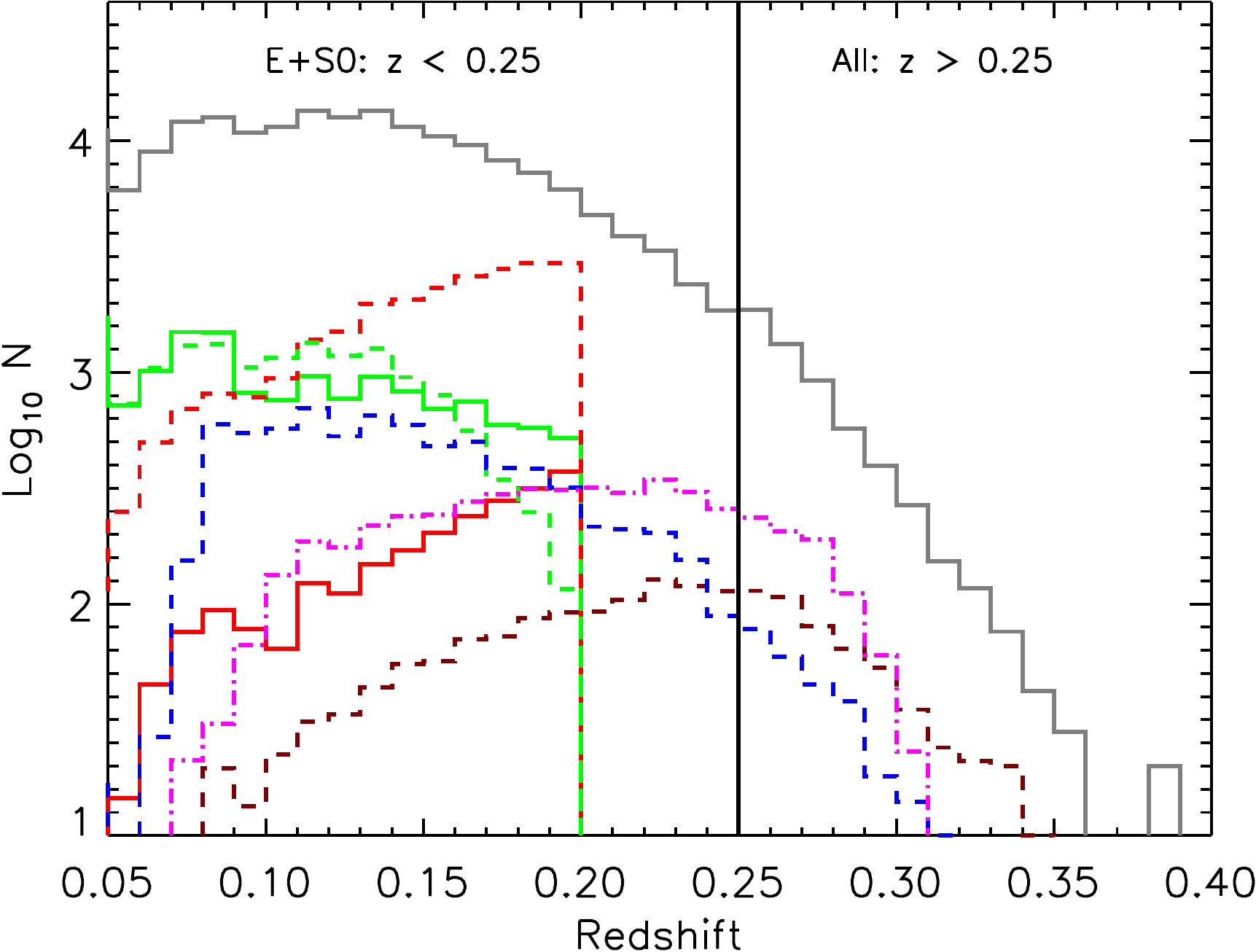}
  
  \vspace{0.5cm}
  \includegraphics[scale = .42]{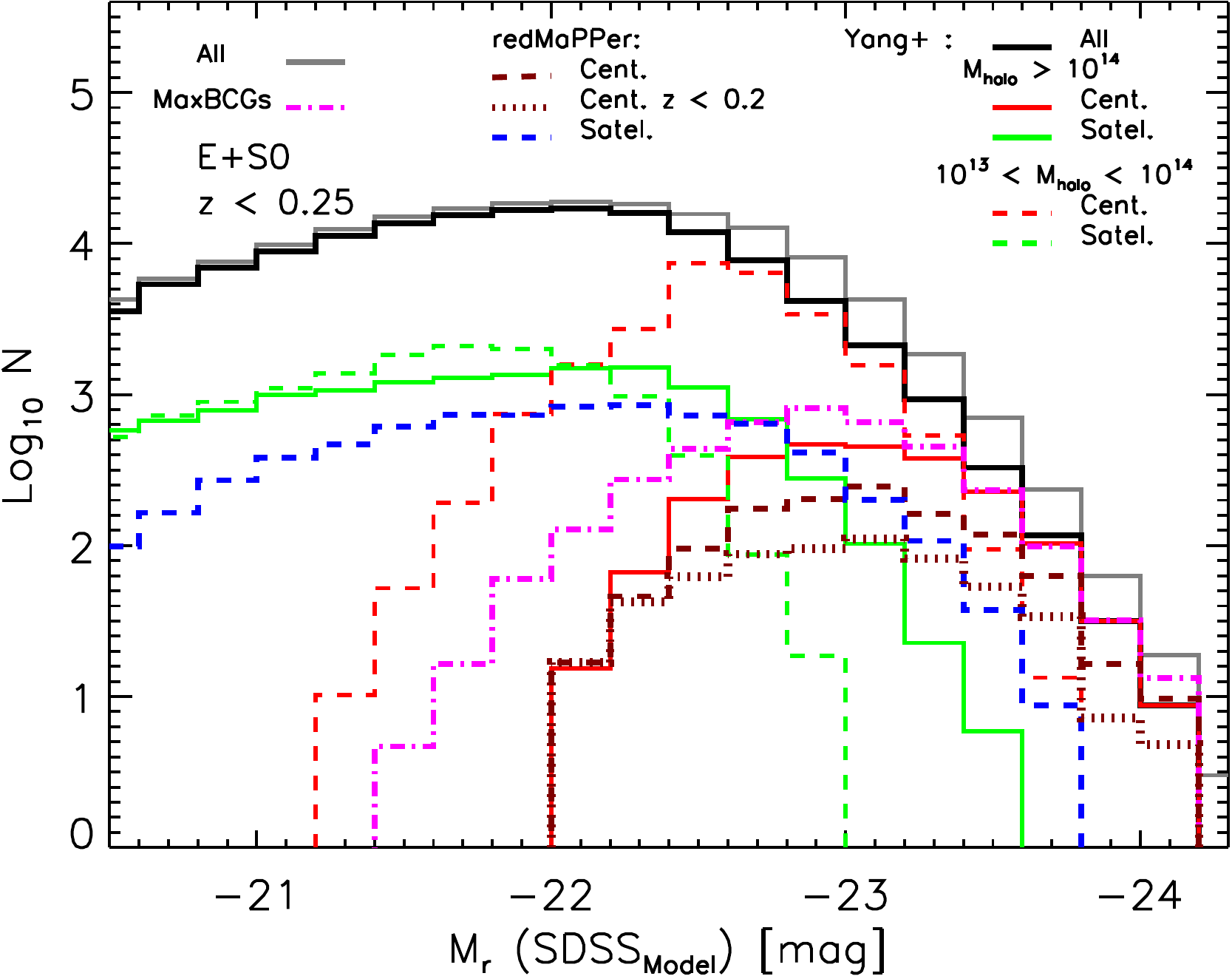}
  \caption{Redshift and luminosity distribution of objects in the group catalogs we use.  Grey histograms in the two panels show the distribution in the full sample.  Other linestyles are as indicated in legend in the bottom panel.  At $z\le 0.25$, objects were weighted by the probability of being an E+S0. Only $z\le 0.25$ objects were used to make the histograms in the bottom panel. Note that the {\tt Yang+} catalog includes only galaxies at $z \le 0.2$; restricting the full sample (grey histogram) to galaxies with $z\le 0.2$ gives the same distribution as for the full {\tt Yang+} catalog (black histogram). Note that the vast majority of objects having $-23.4\le M_r\le -22.4$ are centrals in groups less massive than $10^{14}M_\odot$.  Using the {\tt Yang+SerExp} catalog (described in Section~\ref{yang+pymorph}) and plotting versus {\tt SerExp} magnitudes instead mainly shifts the counts to brighter magnitudes, but the differences between the samples remain.
  }
 \label{dndz}
\end{figure}

\begin{figure}
 \centering
 \includegraphics[scale = .42]{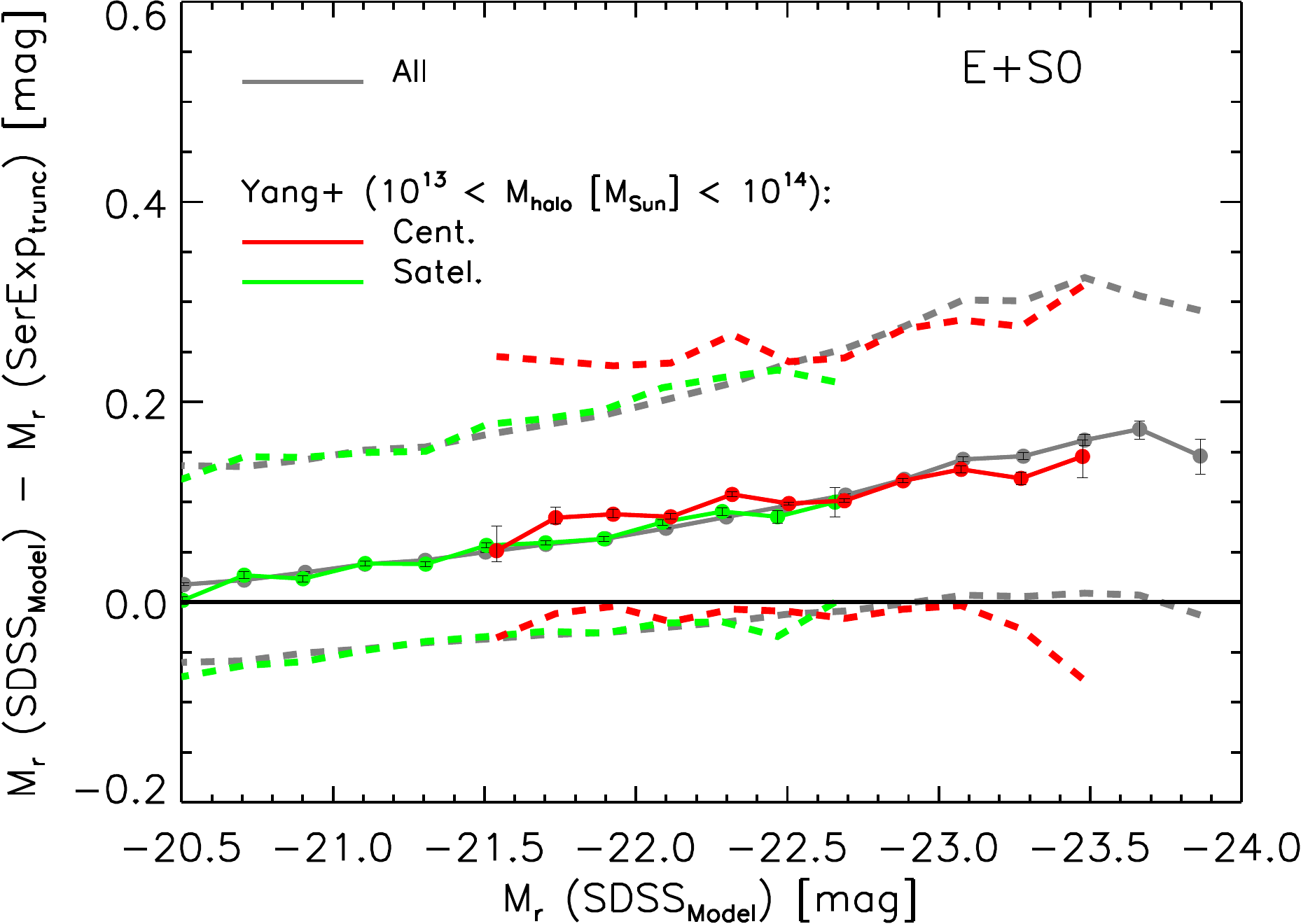}
 \caption{Difference between {\tt Model} and {\tt SerExp} (truncated) magnitudes for galaxies in {\tt Yang+} having group masses between $10^{13}M_\odot$ and $10^{14}M_\odot$, as a function of {\tt Model} magnitudes.  The median difference defined by all the E+S0 galaxies (grey; same as corresponding curve in top panel of Figure~\ref{whichMag}) is significantly different from zero and is almost exactly the same for centrals (red) as for satellites (green). }
 \label{lowMhalo}
\end{figure}

Red and green symbols show objects {\tt Yang+} classify as centrals and satellites, respectively.  Notice that the centrals define a rather tight $M_{\rm Halo}$-{\tt Model} magnitude correlation which has a sharp boundary.  This is because of the {\tt Yang+} definition of $M_{\rm Halo}$, and will be important later (see discussion related to Figures~\ref{lowMhalo} and \ref{lowMhalo2}, and Appendix~\ref{secbias}).  More important in the present context are the brown symbols, which show the centrals in {\tt redMaPPer} (the {\tt redMaPPer} satellites are all the points which are not brown).  Below $M_{\rm Halo}\sim 10^{14}M_\odot$ only a small fraction of the {\tt Yang+} centrals are also centrals in {\tt redMaPPer} (i.e., {\tt redMaPPer} labels most {\tt Yang+} centrals as satellites -- remember that this is the subsample of galaxies which is common to both group catalogs).  If {\tt Yang+} are correct, then {\tt redMaPPer} is wrongly linking together objects which are really in separate halos -- this will compromise efforts to use {\tt redMaPPer} to address assembly bias like effects.  On the other hand, if {\tt redMaPPer} is correct, then the {\tt Yang+} misclassifications will make centrals and satellites seem more similar than they really are.  This disagreement between the two groups complicates any attempt to draw unambiguous conclusions about the central-satellite difference in lower mass halos.  

In view of the central-satellite disagreement, in what follows, we work primarily with the {\tt Yang+} objects which {\tt Yang+} classified as being in halos more massive than $10^{14}M_\odot$.  While Figure~\ref{yangRedmapper} suggests this is a reasonable choice, things are not completely straightforward, since only about one in three of the {\tt Yang+} centrals is in the {\tt redMaPPer} catalog.  We show this in Figure~\ref{dndz}, where $N$ is the total number of objects in each catalog (i.e. not the small subset which were common to both catalogs) weighted by $p$(E+S0).  (Since we only have BAC weights at $z\le 0.25$, we set $p$(E+S0)=1 for all objects at $z > 0.25$, since this is very likely to be realistic. We have checked that making cuts on concentration index or color smooths out the small plateau at $z=0.25$ in the grey histogram, but makes essentially no difference to the others.)  The bottom panel only includes galaxies at $z\le 0.25$. Remember that the {\tt Yang+} catalog only includes objects a $z < 0.2$. Restricting the full sample (grey histogram) to galaxies with $z\le 0.2$ gives the same distribution as for the full {\tt Yang+} catalog (black histogram). For completeness, the bottom panel also shows the subsample of the centrals in the {\tt redMaPPer} catalog at $z < 0.2$ (brown dotted line). (We show the number of objects $N$, rather than the comoving number density, since our goal is to show how many objects contribute to each bin in the Figures which follow.)

Note that, in the bottom panel, there are $3\times$ more centrals in {\tt Yang+} than in {\tt redMaPPer}, even when restricting {\tt Yang+} to $M_{\rm Halo}\ge 10^{14}M_\odot$.  This remains true if we use the {\tt Yang+SerExp} catalog (described in Section~\ref{yang+pymorph}) and show the counts as a function of {\tt SerExp} magnitudes instead (the main difference is a shift to brighter magnitudes).  We assume that the reason for this  difference (i.e. the factor of $3\times$) is not the central-satellite designation, but the group richness:  presumably many of the {\tt Yang+} objects are centrals of lower richness groups which failed the {\tt redMaPPer} richness cut.  In partial support of this, we note that the number of {\tt redMaPPer} satellites is nearly the same as {\tt Yang+}, despite having $3\times$ fewer centrals; this is consistent with them being more massive.  In what follows, we use all the {\tt redMaPPer} objects within $z\le 0.25$.  We do not include higher $z$ objects because we do not have E+S0 classifications above $z=0.25$.  (We have checked, but do not show here, that including the red and/or high-concentration objects from $z\ge 0.25$ makes no difference to our results, other than to improve the statistical significance.)  Moreover, {\tt Yang+} only extends to $z=0.2$, and we did not want questions of evolution being different in the two samples to complicate our results.  

Finally, the magenta curves show the distribution of $\sim 5500$ E+S0 central galaxies in the {\tt MaxBCG} catalog of Rykoff et al. (2012) restricted to $z\le 0.25$ and which have PyMorph reductions from Meert et al. (2015).  At the highest luminosities, their comoving density is greater than {\tt redMaPPer} but smaller than the $\ge 10^{14}M_\odot$ {\tt Yang+} centrals, so we expect them to be intermediate in mass as well.  This is our primary reason for including this sample.

\subsection{Combining {\tt PyMorph} with {\tt Yang+}}\label{yang+pymorph}
We remarked in the previous subsection that {\tt Model} photometry plays an important role in the {\tt Yang+} catalog.  This raises the question of how the catalog is modified if we use {\tt PyMorph} photometry instead.  Although the appropriate thing to do is to re-run the algorithm, this is well-beyond the scope of the current study.  Instead, we have performed the following simple but reasonable procedure.

For each {\tt Yang+} group, we assume that although changing the photometry may change the central-satellite designation within a group, it will not change group membership.  We then replace the {\tt Model} photometry with {\tt PyMorph} values for each group member, and sum these new values to obtain a new estimate of the total group luminosity.  We rank-order this quantity.  Since this new ordering is different from that based on {\tt Model} photometry, we reassign halo masses based on this new rank ordering.  We also {\em define} the most luminous galaxy in a group as the central.  Doing this separately for {\tt SerExp} and {\tt Ser} photometry gives us {\tt Yang+SerExp} and {\tt Yang+Ser} catalogs.  
For both catalogs, the set of halo masses associated with the original {\tt Yang+} catalog is unchanged (by construction), but the mapping between $M_{\rm Halo}$ and central galaxy luminosity is modified.

The main effect to a plot like Figure~\ref{yangRedmapper} is to shift the points slightly to brighter magnitudes -- we explore the consequences of this for Halo Model like analyses elsewhere -- but the motivation for splitting the sample at $M_{\rm Halo}\sim 10^{14}M_\odot$ remains.  Since the shifts are small, we have not included a plot showing, e.g., $M_{\rm Halo}$ versus {\tt SerExp} magnitude in the {\tt Yang+SerExp} catalog.  However, defining the {\tt Yang+SerExp} and {\tt Yang+Ser} catalogs is important for what follows (see also Appendix~\ref{secbias}).  

\section{Comparison of centrals and satellites over a range of galaxy and halo masses}

\subsection{Centrals and satellites in poor groups}
Figure~\ref{lowMhalo} shows the difference between {\tt SerExp} and {\tt Model} photometry, as a function of {\tt Model} magnitude.  (The {\tt SerExp} magnitudes are based on truncating the fits at $7.5a_e$ as described previously.)  Grey symbols show the full E+S0 sample; red and green curves show the subset of galaxies identified as being centrals (red) and satellites (green) in halos having masses between $10^{13}M_\odot$ and $10^{14}M_\odot$ in the original {\tt Yang+} catalog.  They are both very similar to the grey curve, which shows the median trend for all E+S0 galaxies (from top panel of Figure~\ref{whichMag}).  Hence, either the {\tt Yang+} central/satellite designations are completely random, or the difference between {\tt Model} and {\tt SerExp} magnitudes does not depend on whether a galaxy is a central or a satellite.  If the latter, then either these groups are too low mass to have a significant ICL component, or what ICL is present does not play an important role in the PyMorph-SDSS difference. In either case, the PyMorph-SDSS difference, at least for galaxies in groups less massive than $10^{14}M_\odot$, is real -- it is not just semantics.

\begin{figure}
  \centering
  \includegraphics[scale = .42]{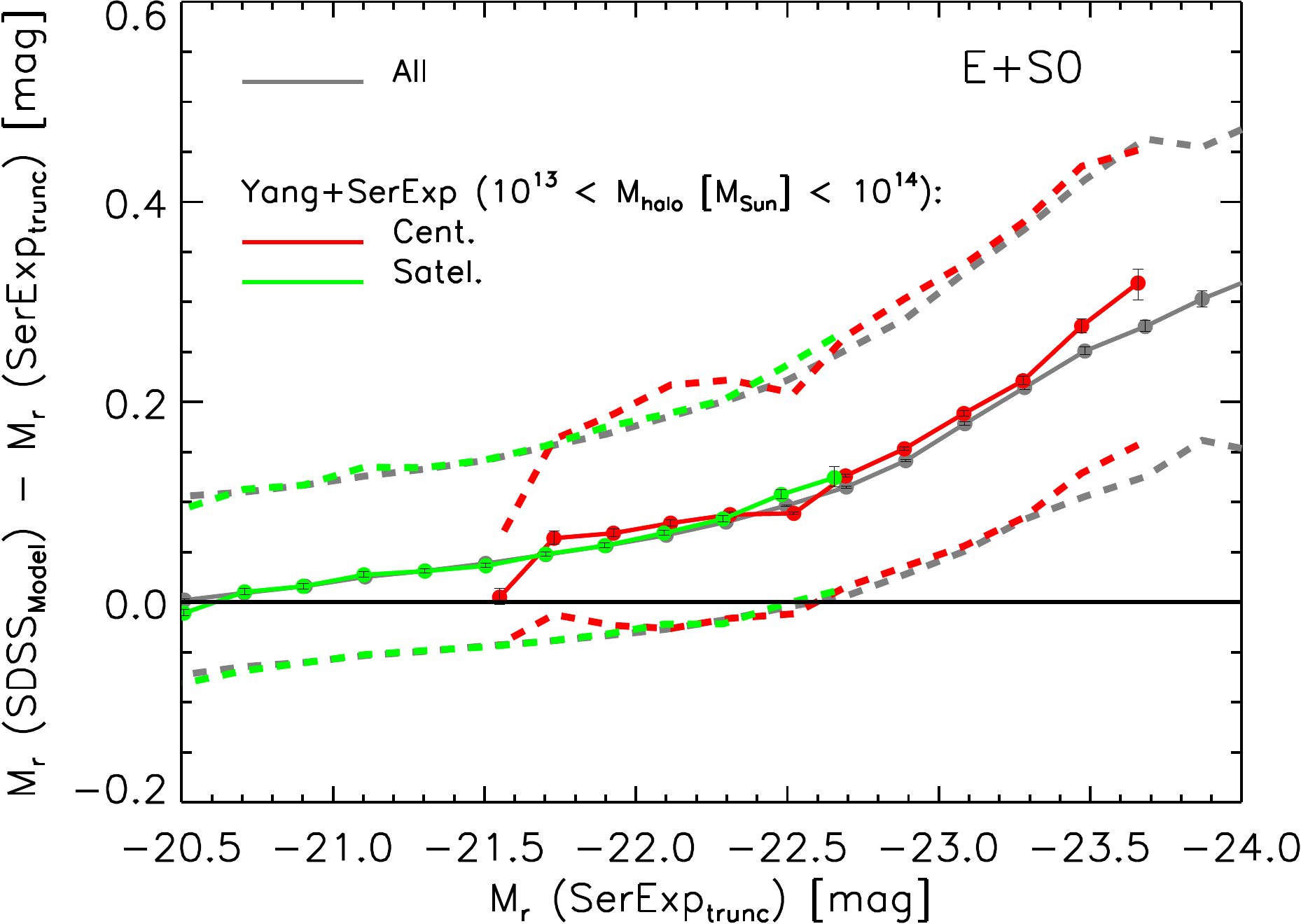}
  \caption{Same as previous Figure, but now shown as a function of {\tt SerExp} magnitude in the {\tt Yang+SerExp} catalog.  As in the previous figure, centrals and satellites are similar to the bulk of the population, even though the median difference is larger than in Figure~\ref{lowMhalo}.}
 \label{lowMhalo2}
\end{figure}

Although Figure~\ref{whichMag} indicates that the difference depends strongly on which quantity is used as reference, we show in Appendix~\ref{secbias} that simply plotting versus {\tt SerExp} magnitudes (instead of {\tt Model}) is biased by a selection effect which arises because $M_{\rm Halo}$ in the {\tt Yang+} group catalog is tightly correlated with {\tt Model} magnitude (c.f. Figure~\ref{yangRedmapper}), but less so with {\tt SerExp}.  For this reason, Figure~\ref{lowMhalo2} shows the difference in the {\tt Yang+SerExp} catalog, where $M_{\rm Halo}$ is correlated with {\tt SerEXp} rather than {\tt Model}.  (For a given group, the central-satellite designation in {\tt Yang+SerExp} may be different from that in {\tt Yang+}.  Defining the `central' as the `brightest' is important:  using the {\tt Yang+} designation here produces a noticable bias, see Figure~\ref{lowMhaloBias}.)  Note in particular that although the median difference (grey) is larger than in Figure~\ref{lowMhalo}, centrals and satellites are still remarkably similar.  Hence, the agreement of the red and green lines with the grey one in Figures~\ref{lowMhalo} and Figure~\ref{lowMhalo2} strongly suggests that centrals and satellites are similar, and that the PyMorph-SDSS difference is not just semantics.

In this context it is interesting to note that Luminous Red Galaxies (LRGs) are believed to populate groups having masses of a few times $10^{13}M_\odot$.  This is similar to the mass scale we are considering here.  For LRGs at $z\sim 0.34$, a stacking analysis shows that the characterisic scale where the ICL becomes apparent is $\sim 100$~kpc \cite{lrgICL}, where the surface brightness is $28$~mags~arcsec$^{-2}$.  Accounting for surface brightness dimming between $z\sim 0.15$ and $z=0.34$ would make this $27.4$~mags~arcsec$^{-2}$.  This is substantially fainter than the $\sim 26$ mag/arcsec$^2$ which corresponds to $\sim 1\%$ of the sky value in the individual $r$-band SDSS images of the objects which contribute to Figures~\ref{lowMhalo} and \ref{lowMhalo2}.  I.e., the individual images will show no sign of the ICL, so it is very unlikely that PyMorph's Sersic-based fits are sensitive to it.  And indeed, Figures~\ref{prof1} and \ref{prof2} show that the PyMorph fits are rather good within $R_{1\%{\rm sky}}$.  Stated differently, the ICL component in the LRG stacks shows up as excess light compared to a Sersic profile (with $n=5.5$) on scales larger than $\sim 8R_e$, but Figures~\ref{prof1} and \ref{prof2} show that the individual images typically have $R_{1\%{\rm sky}}\approx 4R_e$, which is why they show no sign of the ICL.

Finally, note that a de Vaucouleurs profile is not a good fit to the regions within $\sim 8R_e$, neither for LRGs nor for the galaxies we are considering here.  Thus, if one thinks of a Sersic profile as describing light in excess of a de Vaucouleurs profile, then this excess light is {\em not} ICL.  The ICL shows up as an additional departure from a Sersic profile beyond $8R_e$.  This is consistent with our assertion above:  at least for galaxies in groups less massive than $10^{14}M_\odot$, the PyMorph-SDSS difference is not due to the ICL.

\begin{figure*}
 \centering
 \includegraphics[scale = .42]{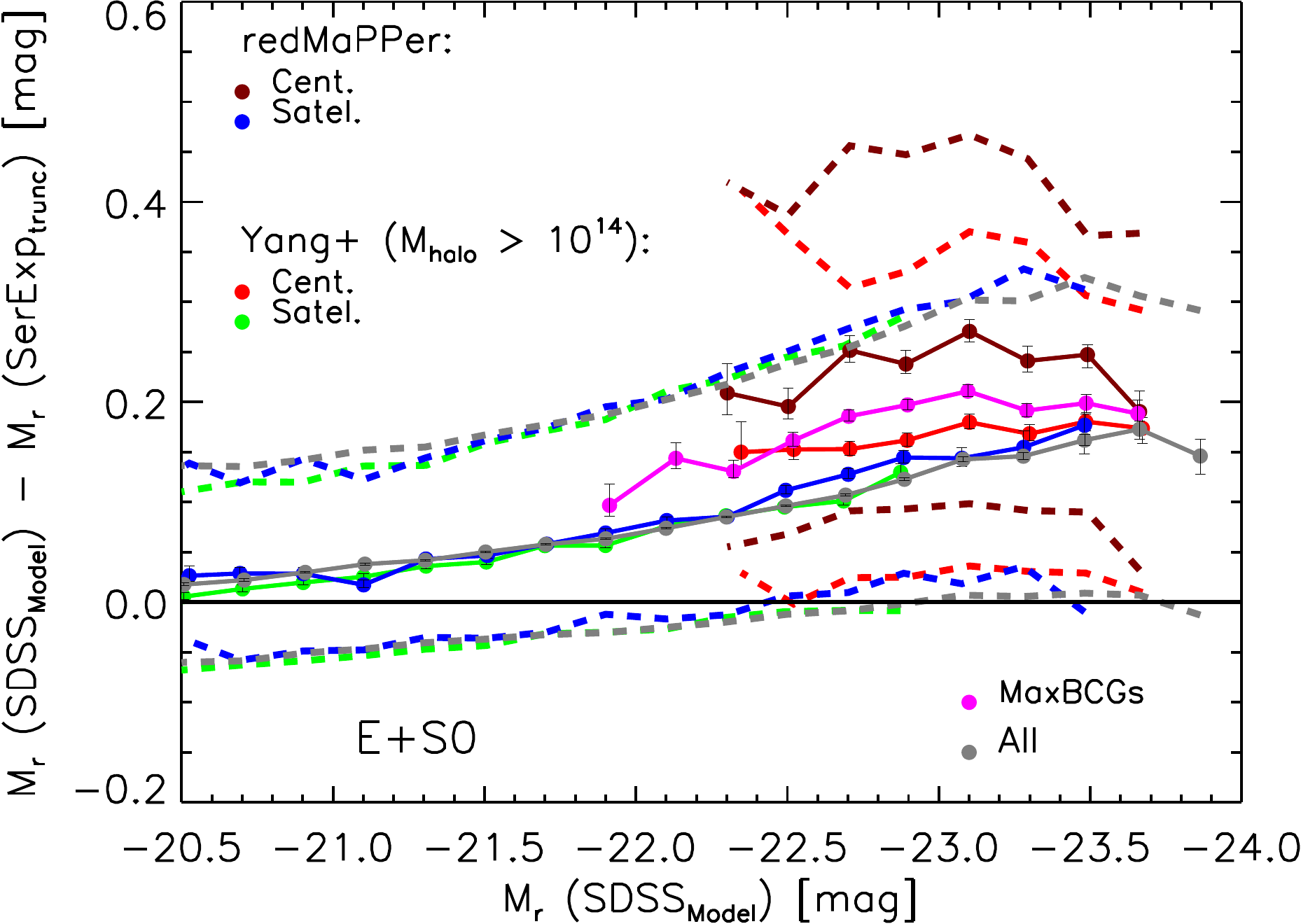}
 \includegraphics[scale = .42]{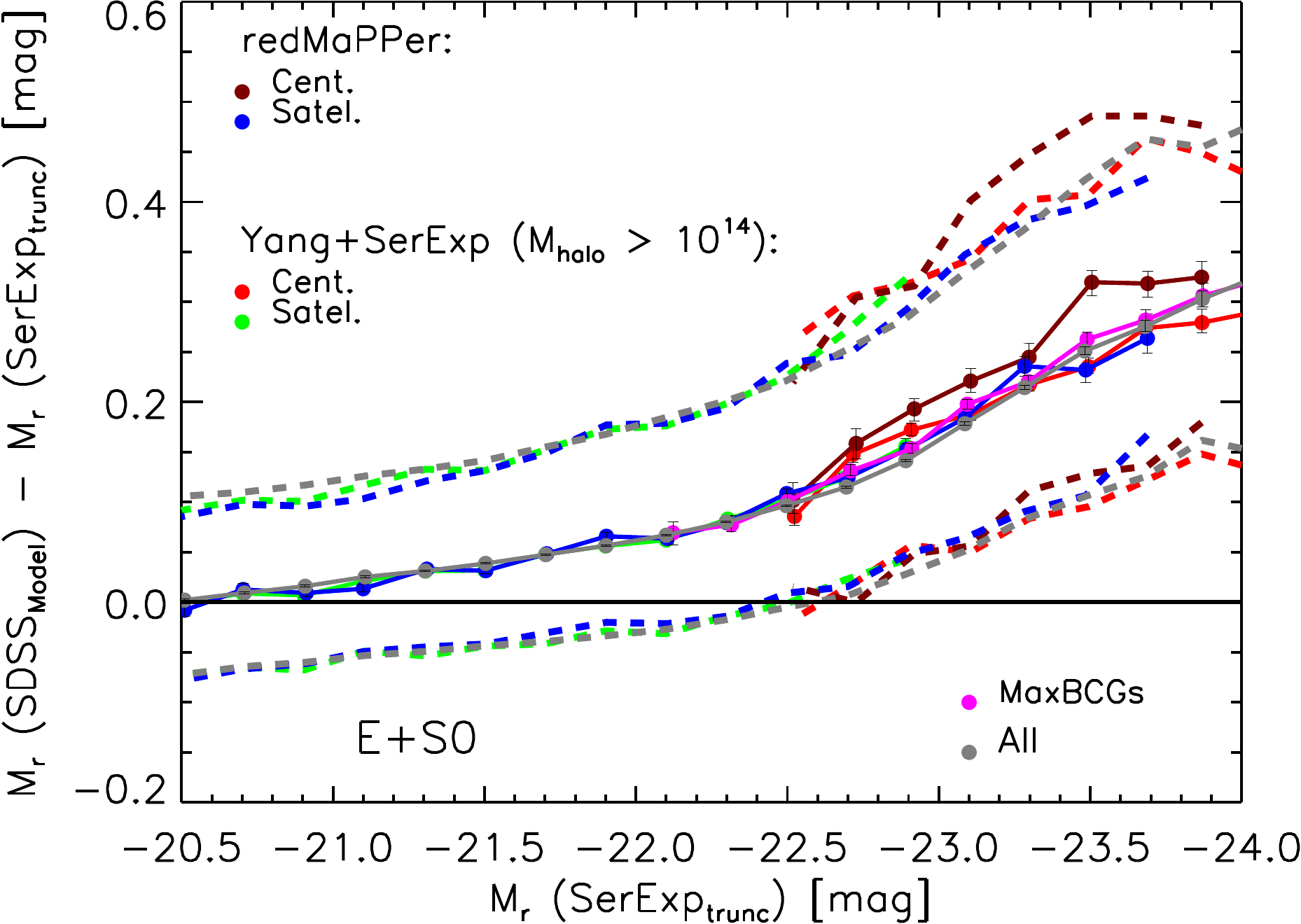}
 \caption{Difference between (truncated) {\tt SerExp} and {\tt Model} photometry, as a function of {\tt Model} (left) and {\tt SerExp} magnitude (right).  Grey symbols show the full E+S0 sample; red and green symbols show the subset of these E+S0s which are centrals and satellites in groups whose halo mass is greater than $10^{14}M_\odot$ in the {\tt Yang+} (left) and {\tt Yang+SerExp} (right) catalogs; brown and blue symbols show correponding measurements in the {\tt redMaPPer} catalog; and magenta symbols show the {\tt MaxBCG} centrals.  In the left panel, satellites are similar to the full sample, whereas centrals tend to be brighter.  However, differences between centrals and satellites are substantially smaller than the differences between {\tt SerExp} and {\tt Model} photometry except for the most massive (i.e. {\tt redMaPPer}) groups.  In the right panel, centrals and satellites are more similar even though the median offset from zero is larger.}
 \label{Ldiffa}
\end{figure*}

Now, the PyMorph-SDSS difference for the full sample of galaxies (grey curve) is similar to that for these centrals and satellites.  Since the vast majority of these are in lower mass halos, ICL effects do not play a role for the vast majority of objects which contribute to the grey curve.  These are the objects which Bernardi et al. (2013, 2016, 2017) used to estimate the $z\sim 0.1$ luminosity and stellar mass functions.  Hence, the difference between these mass functions and those based on SDSS pipeline photometry are real:  it is incorrect to attribute it to the semantics of whether or not one includes the ICL when estimating the light from a galaxy.  

\subsection{Centrals and satellites in massive clusters}

We now consider galaxies in more massive groups.  
Figure~\ref{Ldiffa} shows the difference between {\tt SerExp} and {\tt Model} photometry, as a function of {\tt Model} (left) and {\tt SerExp} magnitude (right).  The grey curves in the two panels are for the full sample and are taken from Figure~\ref{whichMag} as before.  Red and green symbols and curves show the subset of galaxies which {\tt Yang+} identify as being centrals and satellites in halos more massive than $10^{14}M_\odot$ -- these are shown only on the left panel due to the selection effect just mentioned above (and discussed in Appendix~\ref{secbias}).  In the right panel, the red and green symbols and curves show the corresponding subset of galaxies in our {\tt Yang+SerExp} sample.  (I.e., they are in halos more massive than $10^{14}M_\odot$, but because the assignment of halos masses -- and in some cases the central satellite designation -- has changed, they are not exactly the same objects as in the panel on the left.)  Brown and blue symbols and curves in both panels show centrals and satellites in the {\tt redMaPPer} catalog, and magenta shows the {\tt MaxBCG} centrals.

In both panels, the satellites are in good agreement with the grey:  satellites are similar to the average over the full population, even at $M_r<-23$.  In the panel on the right, centrals and satellites are also in good agreement (only ther {\tt redMapper} centrals are slightly offset), even though the median differences between SDSS and {\tt SerExp} are larger compared to the panel on the left.  Only in the panel on the left do the centrals tend to be slightly more luminous, with the difference increasing with group mass (recall that the {\tt redMaPPer} groups are more massive than {\tt Yang+}).  At $M_r<-23$, where the grey curve indicates the average difference is 0.15~mag, the {\tt redMaPPer} centrals show an additional 0.08~mag difference.  Clearly, the centrals of the most massive clusters, which are less than 30\% of the rarest most luminous objects (Figure~\ref{dndz}), are different from the vast majority of the galaxy population.

However, even for these centrals, it is not obvious that one can attribute the additional offset entirely to ICL-related effects for the reasons given in Section~\ref{images}.
As Figures~\ref{prof1}--\ref{prof3} illustrate (Figure~\ref{prof3} is a {\tt redMaPPeR} central), the PyMorph fits are usually accurate out to $\sim 1\%$ of sky.  If there is a difference beyond this scale, it is in the sense that the ICL will be yet another addition to the {\tt SerExp} estimate.  

\subsection{Differences in sky and profile shape}
When discussing Figure~\ref{whichMag} we noted that the median trend for the full E+S0 sample (labelled `All') is due to two effects:  one is due to differences in the PyMorph and SDSS sky estimates, and the other to the increased freedom which the two component {\tt SerExp} profile has compared to {\tt Model}, which, for E+S0s, is basically a single de Vaucouleurs profile.  This raises the question of which effect dominates the central-satellite differences we see in Figure~\ref{Ldiffa}?  

\begin{figure}
 \centering
 \includegraphics[scale = .42]{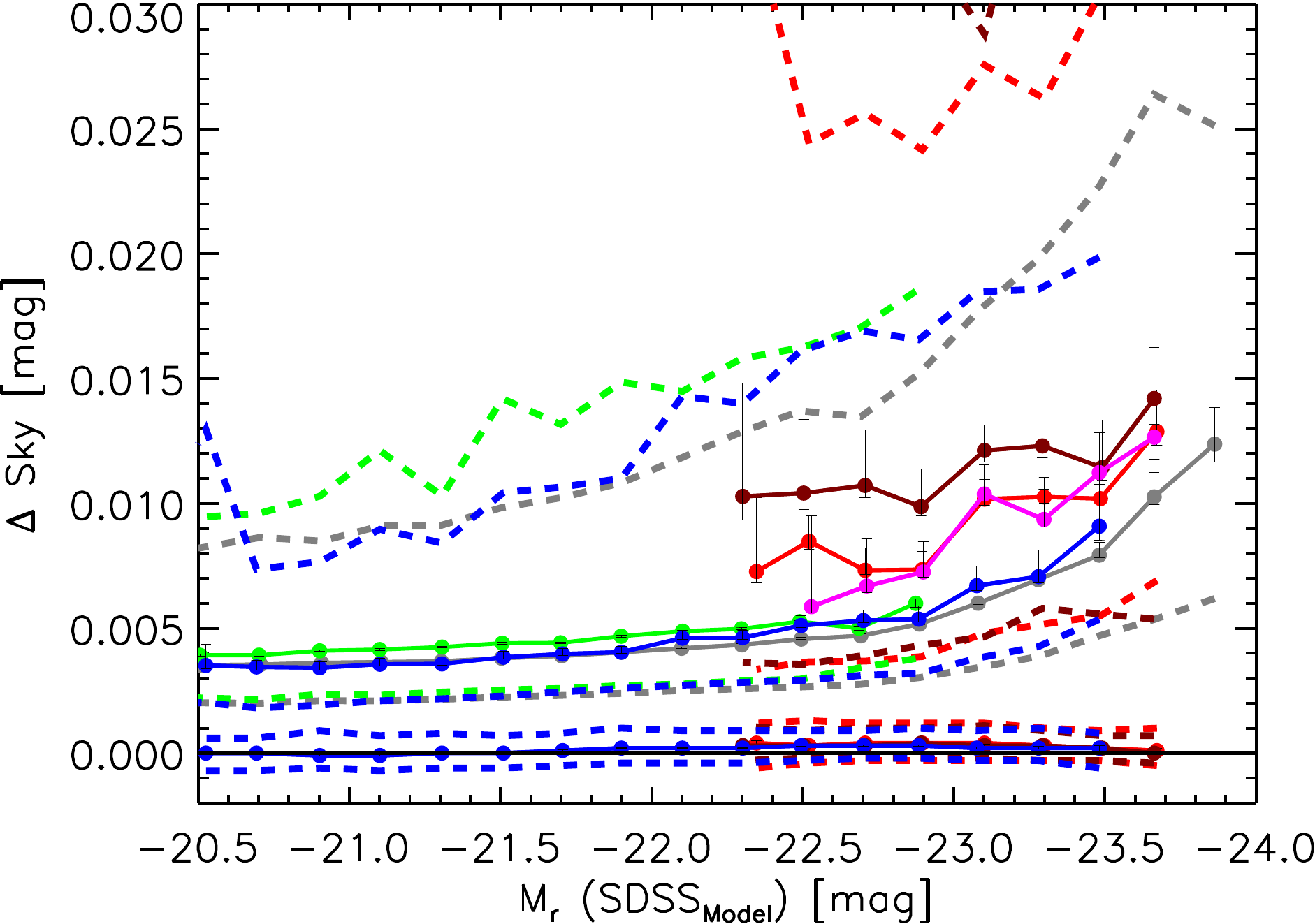}
 \caption{Comparison of sky estimates; colors are same as Figure~\ref{Ldiffa}.  Top set of curves, showing larger offset and asymmetric scatter, show difference between PyMorph {\tt SerExp} and SDSS DR7 sky estimates ($\Delta$ Sky $=$ Sky (PyMorph {\tt SerExp}) - Sky (SDSS)).  The SDSS sky estimate is brighter; this overestimate is largest for the most luminous objects, and is even larger for centrals than for satellites.  Bottom set of curves, showing no offset and smaller scatter, show difference between PyMorph {\tt SerExp} and PyMorph {\tt deV} sky estimates (i.e. $\Delta$ Sky $=$ Sky (PyMorph {\tt SerExp}) - Sky (PyMorph {\tt deV})). The difference in these sky estimates is insignificant. This conclusion is similar if we compare PyMorph {\tt Ser} and PyMorph {\tt SerExp} sky estimates.}
 \label{sky}
\end{figure}

To address this, the lower set of curves in Figure~\ref{sky} compare the sky estimates in the PyMorph {\tt deV}, {\tt Ser}, and {\tt SerExp} fits.  These show no significant differences from zero.  This is in marked contrast to the top set of curves showing a large asymmetric scatter.  These indicate that the SDSS DR7 sky estimate is brighter than PyMorph's for the most luminous objects.  Moreover, the SDSS overestimate is even brighter for centrals than for satellites.  The former trend was reported by Fischer et al. (2017, and references therein) who show that the SDSS sky estimate is brighter than both PyMorph and Blanton et al. (2011), while the latter two are in excellent agreement with one another. Fischer et al. argued that it was a consequence of the fact that the SDSS does not use a large enough region from which to estimate the background sky. The issue with the SDSS sky will only be exacerbated in clusters, so the fact that it is worse for centrals than satellites is not unexpected.  (The comparison here uses the SDSS DR7 rather than DR9 sky values, since the PyMorph photometry is based on the DR7 rather than DR9 flux calibrated images.  However, Section~3 and Figures~9 and 10 in Fischer et al. 2017 show that, despite the changes from DR7 to DR9, the bias in the SDSS DR7 sky estimates is also present for the DR9 reductions.)

\begin{figure*}
 \centering
 \includegraphics[scale = .42]{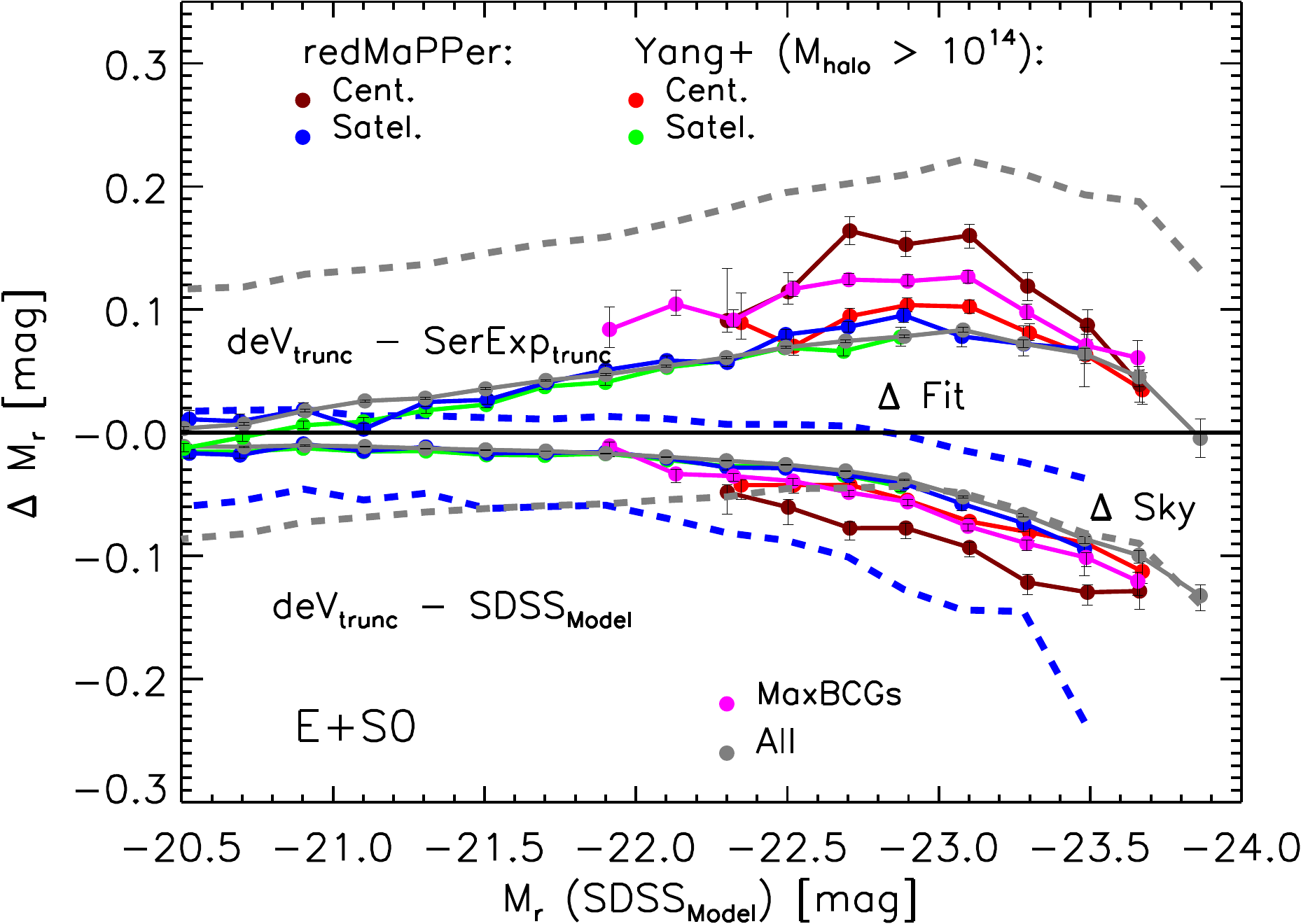} 
 \includegraphics[scale = .42]{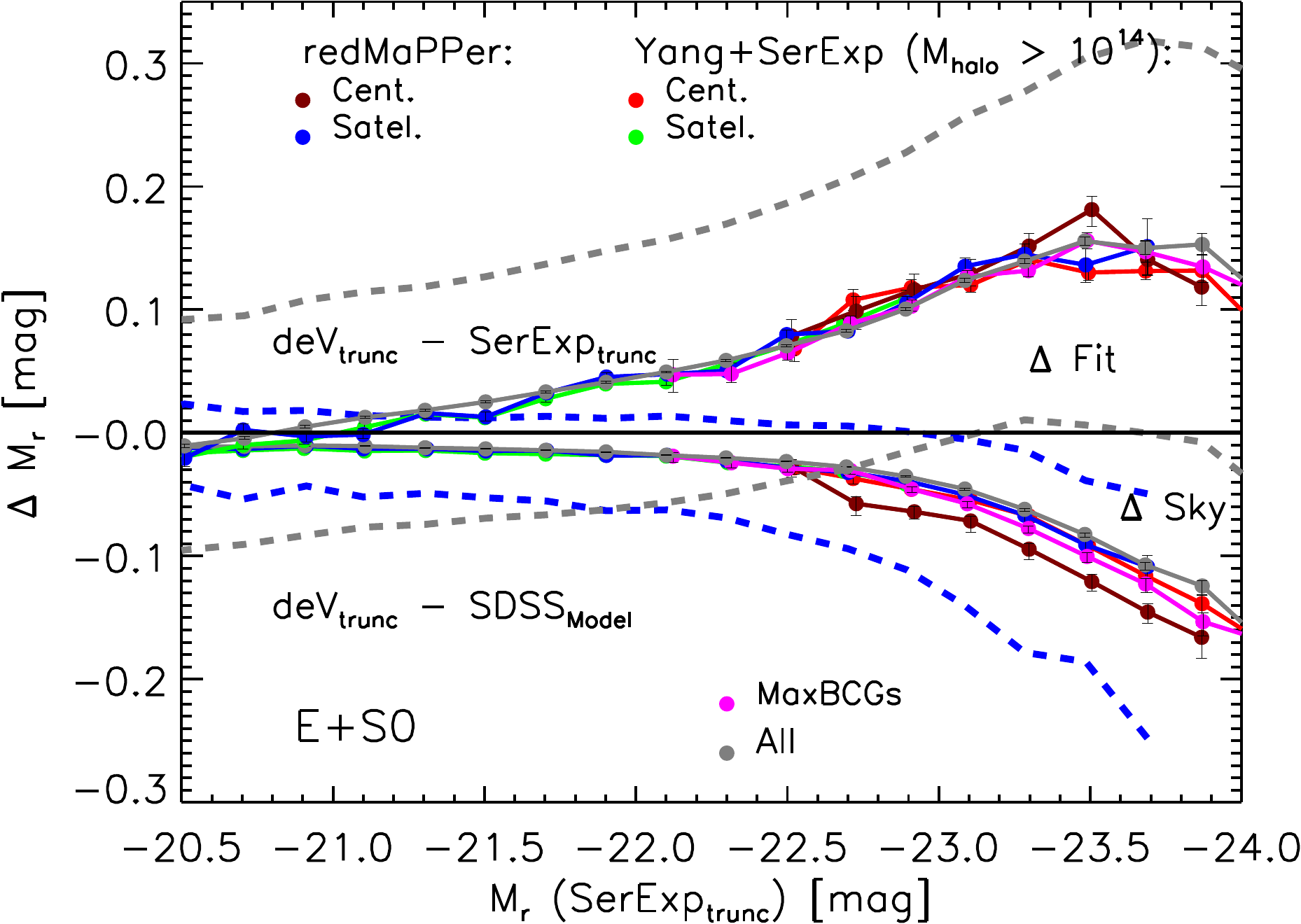} 
 \caption{Same as Figure~\ref{Ldiffa}, but now the difference between {\tt SerExp} and {\tt Model} photometry is broken up into two terms: one isolates sky subtraction effects, and the other is due to true differences in the shape of the light profile.   }
 \label{Ldiffb}
\end{figure*}

\begin{figure*}
 \centering
 \includegraphics[scale = .42]{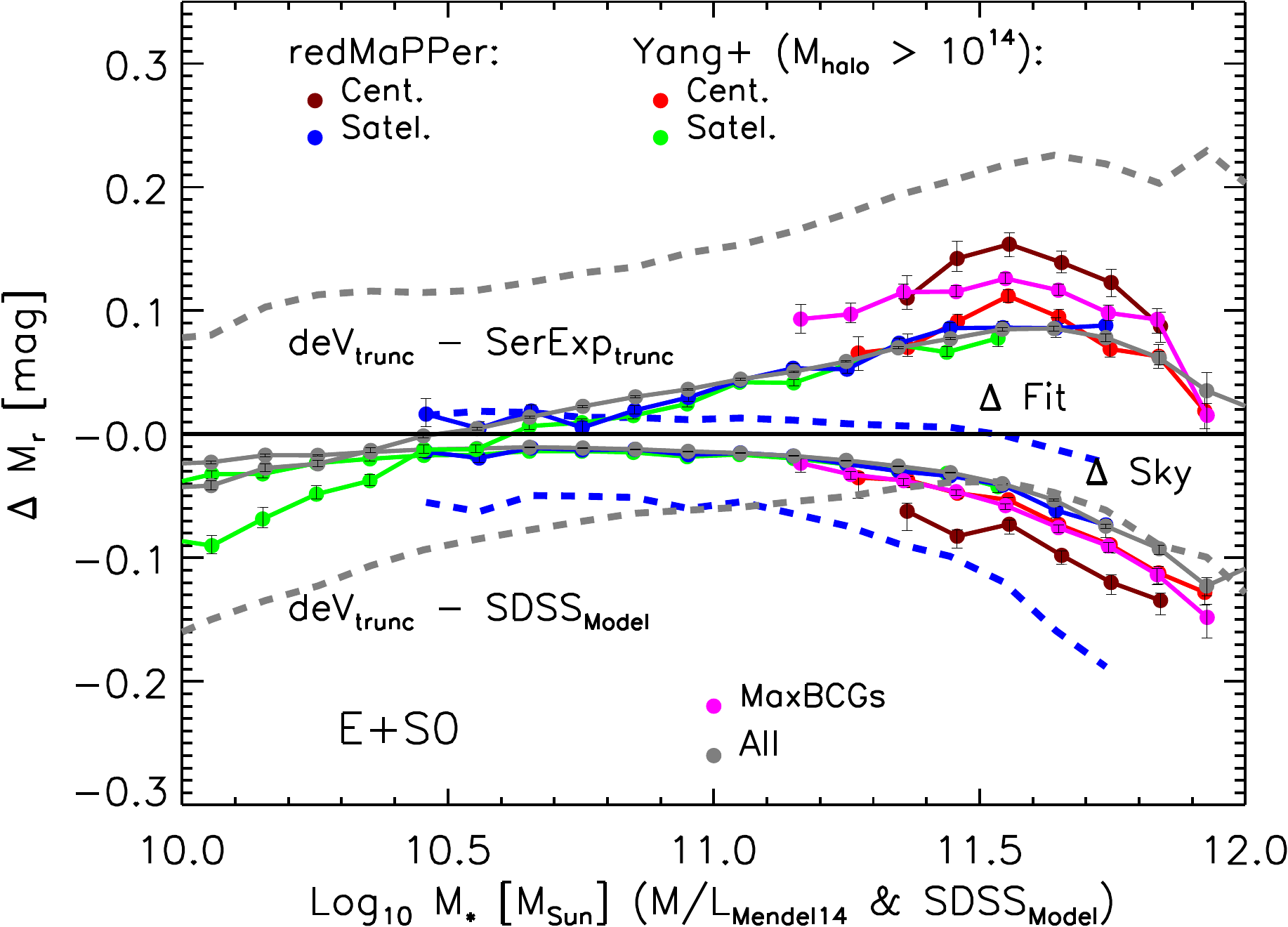}
 \includegraphics[scale = .42]{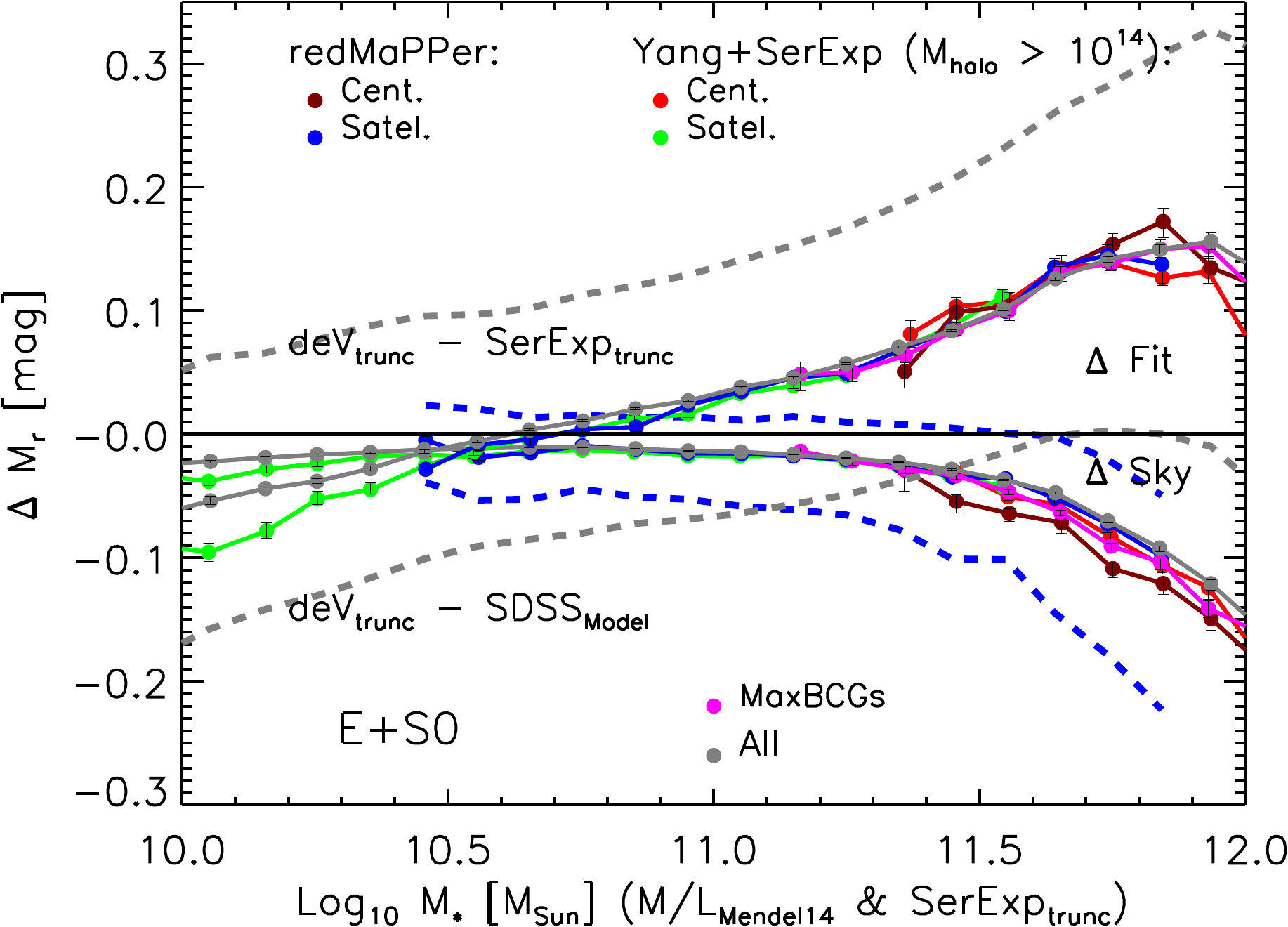}
 \caption{Same as Figure~\ref{Ldiffb}, but now versus $M_*$ rather than absolute magnitude.}
 \label{Mdiff}
\end{figure*}

To see if this can account for all the differences shown in Figure~\ref{Ldiffa} it is useful to write 
\begin{equation}
 {\tt Model}-{\tt SerExp} = ({\tt deV}-{\tt SerExp}) - ({\tt deV}-{\tt Model}),
\end{equation}
where {\tt deV} is the result of using PyMorph to fit a (truncated) de Vaucouleurs profile to the image.  The first term (i.e. {\tt deV}-{\tt SerExp}) isolates the effect of fitting different models whereas the second term (i.e. ${\tt deV}-{\tt Model}$) is entirely due to the difference between the PyMorph and SDSS sky estimates.

Figure~\ref{Ldiffb} shows these two contributions for the various samples shown in Figure~\ref{Ldiffa}.  Comparison with the left hand panel of Figure~\ref{Ldiffa} shows that the sky accounts for a little less than half the total difference, and it affects the {\tt redMaPPer} centrals the most. At $M_r\sim -23$, the choice of profile shape alone accounts for about 0.08~mags when averaged over all E+S0s, with the average for {\tt redMaPPer} centrals being larger by an additional 0.08~mags.  However, in the panel on the right, the central-satellite difference is entirely due to differences in sky.  

Figure~\ref{Ldiffb} makes one more point.  The grey curve in the left panel shows that the ${\tt deV}-{\tt SerExp}$ difference is not monotonic with luminosity.  Section~4 in Fischer et al. (2017) discusses this in more detail and argues that it suggests there are two populations at high luminosities.  However, their analysis did not provide a physically motivated model for the two populations.  Our analysis suggests that the two populations may correspond to centrals and satellites.

Figure~\ref{Mdiff} shows that none of our conclusions are changed if we plot the magnitude difference versus the corresponding stellar mass estimate rather than luminosity.  And Figures~\ref{LdiffSer} and~\ref{MdiffSer} show that they are unchanged if we replace {\tt SerExp} magnitudes with {\tt Ser}. (For all these figures, we use {\tt Yang+} for the panels on the left, and {\tt Yang+Ser} for the panels on the right.)     
In all cases, once the bias in the SDSS sky has been accounted for, the luminosity or stellar mass dependence of the SDSS-PyMorph difference is the same for all satellites, whatever the mass of the cluster they inhabit.  Although for centrals of the most massive clusters, there is an additional effect which increases as cluster mass increases, this additional effect is substantially smaller when shown as a function of Sersic-based rather than SDSS {\tt Model} magnitudes.

To explore the mass dependence further, we divided the {\tt Yang+} and {\tt Yang+SerExp} samples into objects having $14\le \log_{10}(M_{\rm halo}/M_\odot)\le 14.3$ and $\log_{10}(M_{\rm halo}/M_\odot)\ge 14.3$.  We saw a small difference between the two, which is consistent with the mass trend we infer from the comparison with {\tt redMaPPeR}.  Similarly, when {\tt redMaPPer} is split into two subsamples at richness $\lambda=32$, the one with higher richness shows a slightly larger difference. In both cases, the smaller sample size means the statistical significance of the finer mass trend is smaller, so we have not shown these additional trends here.  

We conclude that the analysis in this Section has shown that, for the vast majority of luminous galaxies, the difference between PyMorph and SDSS photometry is not dominated by ICL-like effects.  Rather, it reflects real structural differences, and is {\em not} just a matter of semantics.

\section{Discussion}\label{discus}
The SDSS and PyMorph luminosity estimates differ (Figure~\ref{whichMag}), because the SDSS sky estimate is biased, and because the fitted models differ \cite{F16} . Biases in the SDSS sky estimate are worst for central galaxies in the most massive halos (brown curve in Figure~\ref{sky}).  In general, sky-related biases account for a little less than half the total difference between the SDSS and PyMorph luminosities (lower set of curves in Figures~\ref{Ldiffb}, \ref{Mdiff}, \ref{LdiffSer} and \ref{MdiffSer}). The remaining difference (upper set of curves in Figures~\ref{Ldiffb}, \ref{Mdiff}, \ref{LdiffSer} and \ref{MdiffSer}) is due to fitting different models to the surface brightness profile (i.e. {\tt deV}, {\tt Ser} or {\tt SerExp}).

For the vast majority of galaxies, once biases in the SDSS sky estimate have been accounted for, the SDSS-PyMorph difference does not depend strongly on whether a galaxy is a central or a satellite.
The difference, averaged for the full sample of galaxies (grey curve), is the same as that of central or satellite galaxies of less massive groups (red and green curves in Figures~\ref{lowMhalo} and~\ref{lowMhalo2}, and of satellites in more massive groups (blue curves in Figures~\ref{Ldiffb} and~\ref{Mdiff}). We only see an additional effect for centrals in the most massive halos (which Figure~\ref{dndz} shows are rare), and then only when shown as a function of SDSS rather than Sersic-based photometry. This conclusion is similar whether one uses two-component {\tt SerExp} photometry or single component {\tt Ser} fits (compare Figures~\ref{Ldiffb} and~\ref{Mdiff} with Figures~\ref{LdiffSer} and~\ref{MdiffSer}).

The ICL is expected to be fainter at larger cluster-centric distances or around central galaxies in smaller groups (e.g. Tal \& van Dokkum 2011).  Now, the difference between Sersic-based and SDSS {\tt Model} magnitudes, when averaged over the full population is the same as for centrals or satellites in smaller groups (Figures~\ref{lowMhalo} and \ref{lowMhalo2}).  In fact, Figures~\ref{Ldiffb} and~\ref{Mdiff} show that it is also the same for satellites in more massive groups.  Therefore, we conclude that the vast majority of massive galaxies are not well fit by a simple de Vaucouleurs profile:  the difference from {\tt Model} magnitudes is not due to the ICL, but indicates real structural differences.  Hence, the difference between PyMorph Sersic-based luminosity and stellar mass functions in Bernardi et al. (2013, 2016 and 2017) with respect to estimates based on SDSS {\tt Model} magnitudes is real -- it is not just semantics.

The issue is slightly more complicated for the small number of galaxies which are centrals of the most massive clusters.  Even though the top set of curves in the right hand panels of Figures~\ref{Ldiffb}, \ref{Mdiff}, \ref{LdiffSer} and \ref{MdiffSer} are so similar,
they are offset brightwards in the left panel (compare brown and grey curves).  Hence, there must be differences between these objects and the vast majority of the galaxy population.  However, we believe that even these differences -- which affect less than 30\% of the rarest most luminous objects (Figure~\ref{dndz}) -- should not be attributed to the ICL entirely. This is because the surface brightness where the ICL becomes apparent is fainter than $\sim 27$ mag/arcsec$^2$ (Zibetti et al. 2005; based not on individual galaxies, but a stacking analysis of $\sim 600$ brightest cluster galaxies).  This is fainter than the $\sim 26$ mag/arcsec$^2$ associated with $\sim 1\%$ of the background sky in individual $r-$band SDSS images, so it is very unlikely that the ICL plays a role when fitting to individual images (Figures~\ref{prof1}--\ref{prof3} and related discussion). 

Moreover, we note that the right hand panels of Figure~\ref{Mdiff} and especially Figure~\ref{MdiffSer} show a dramatic change in slope around $2\times 10^{11}M_\odot$.  This is the same mass scale where a number of other scaling relations also change (Bernardi et al. 2011, 2014), and where the galaxy population becomes dominated by slow rotators (Cappellari et al. 2013).  These other features are thought to indicate a change in the typical assembly mechanism of the population.  While this change may also be related to the build-up of the ICL, the features are not caused by it. 

For all these reasons, we believe that PyMorph {\tt SerExp} photometry represents a significant improvement to SDSS-{\tt Model} photometry.  For the vast majority of massive galaxies, the differences between PyMorph and SDSS are real.  In particular, our work shows that previous halo model analyses which used SDSS pipeline photometry when relating the stellar mass of the central galaxy to the dark matter mass of the halo which surrounds it should be redone.  Ascribing the difference compared to analyses based on Sersic magnitudes to the semantics of whether or not one includes the ICL when describing the stellar mass of the central galaxy is incorrect.

We also showed that, when studying correlations at fixed group mass, care must be taken to ensure that one does not mistake selection effects for physical effects.  The {\tt SerExp} and {\tt Model} magnitude difference, when plotted versus {\tt SerExp} magnitude, shows strong trends even though no such trends were apparent when plotting versus {\tt Model} magnitude (compare red and green curves in Figures~\ref{lowMhalo} and~\ref{lowMhaloBias}).  We argued that this is because the group masses in the {\tt Yang+} catalog are tightly correlated with {\tt Model} magnitudes.  A simple model illustrated why selection effects appear when one studies correlations at fixed group mass with a quantity which was not used to define the groups (Appendix~\ref{secbias}).  Indeed, when we reassign halo masses to the {\tt Yang+} groups on the basis of {\tt SerExp} magnitudes, then the selection effect appears when plotting versus {\tt Model} rather than {\tt SerExp} magnitudes (compare Figures~\ref{lowMhalo2} and~\ref{lowMhalo2Bias}).  
This conclusion is general.  For example, we have checked (but do not show here) that the selection effect is even stronger when {\tt Ser} rather than {\tt SerExp} magnitudes are used.
We hope our demonstration of the existence of this sort of pernicious selection effect, our explanation of its cause (Appendix~\ref{secbias}), and how one should correct for it will prevent future confusion.

\section*{Acknowledgements}
We would like to thank the referee for a helpful report.  MB and RKS are grateful to the ICTP for its hospitality during the summer of 2016.

\appendix 

\section{Comparison with single {\tt Ser} photometry}\label{sersic}

\begin{figure*}
  \centering
  \includegraphics[scale = .42]{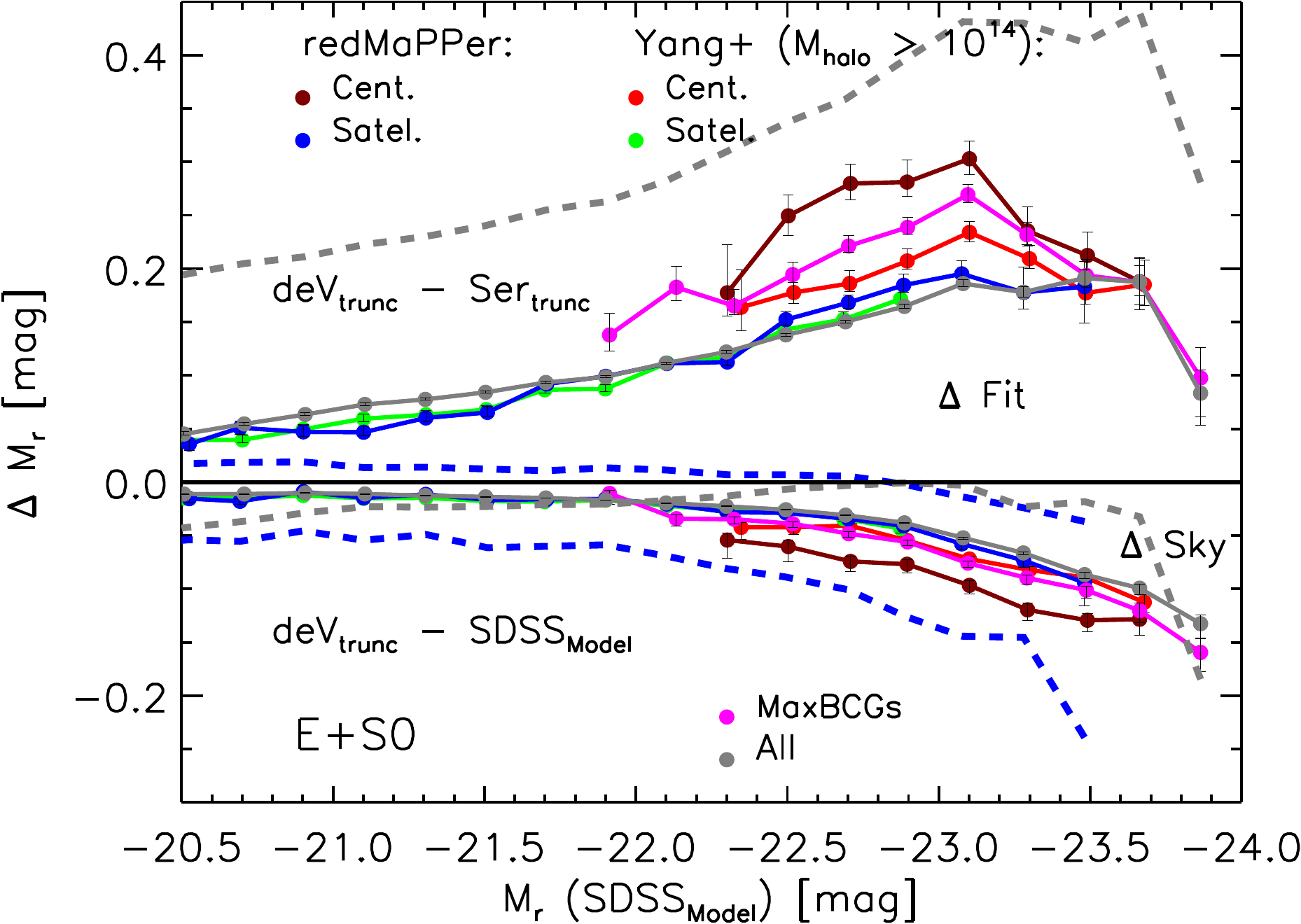}  
  \includegraphics[scale = .42]{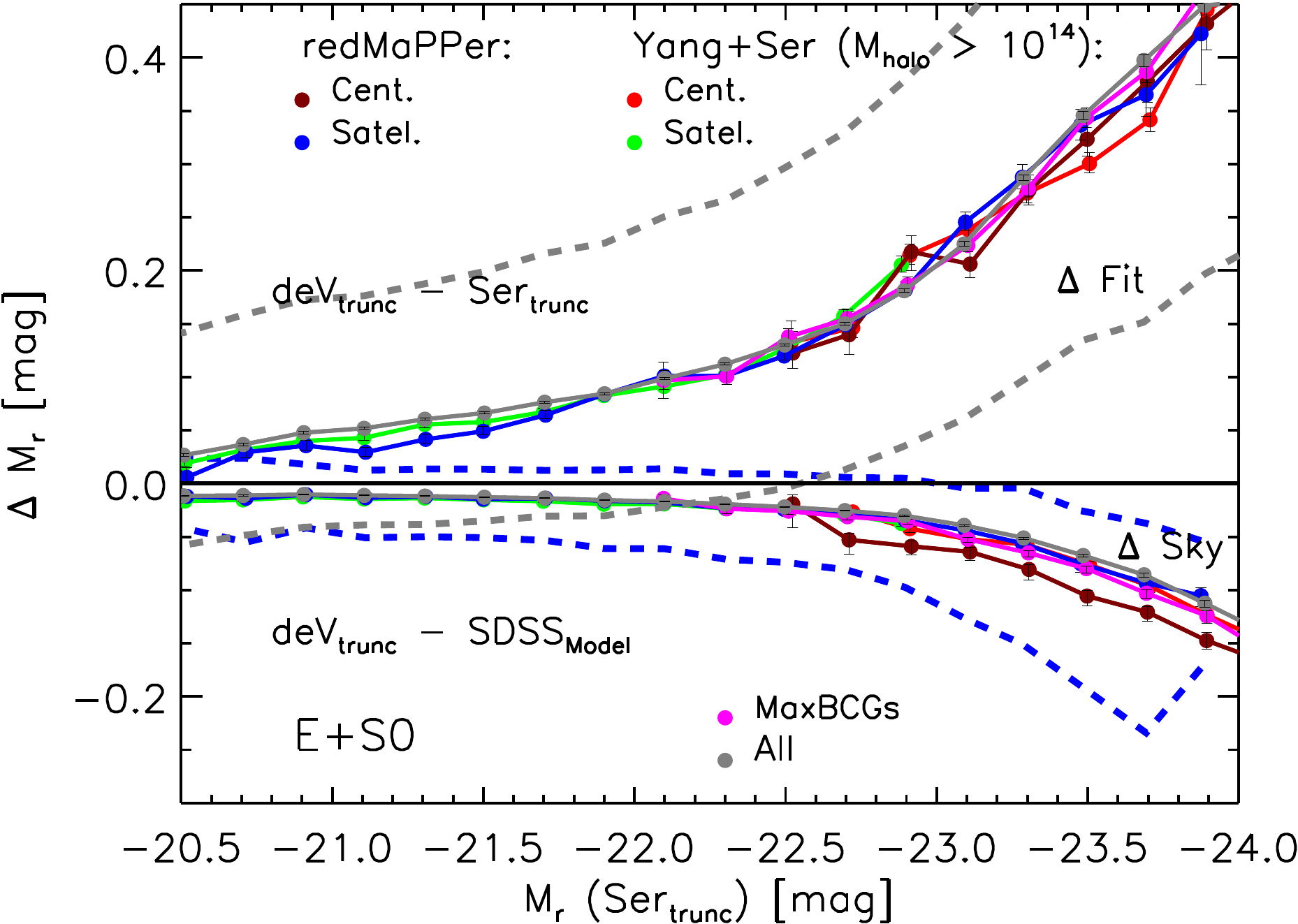} 
 \caption{Same as Figure~\ref{Ldiffb} but with (truncated) {\tt Ser}~mags in place of {\tt SerExp}, and {\tt Yang+Ser} in place of {\tt Yang+SerExp}. }
 \label{LdiffSer}
\end{figure*}

\begin{figure*}
  \centering
 \includegraphics[scale = .42]{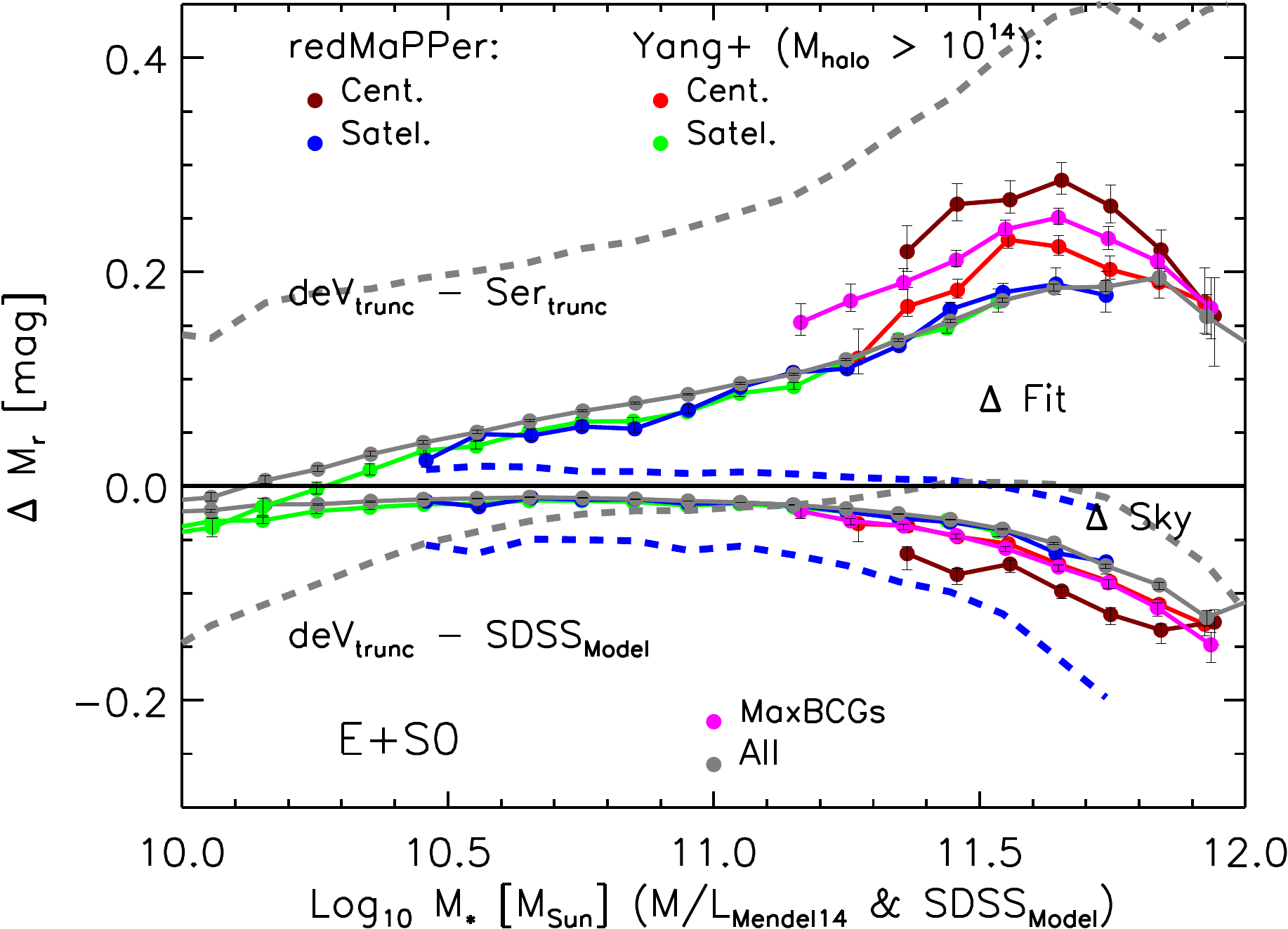}  
 \includegraphics[scale = .42]{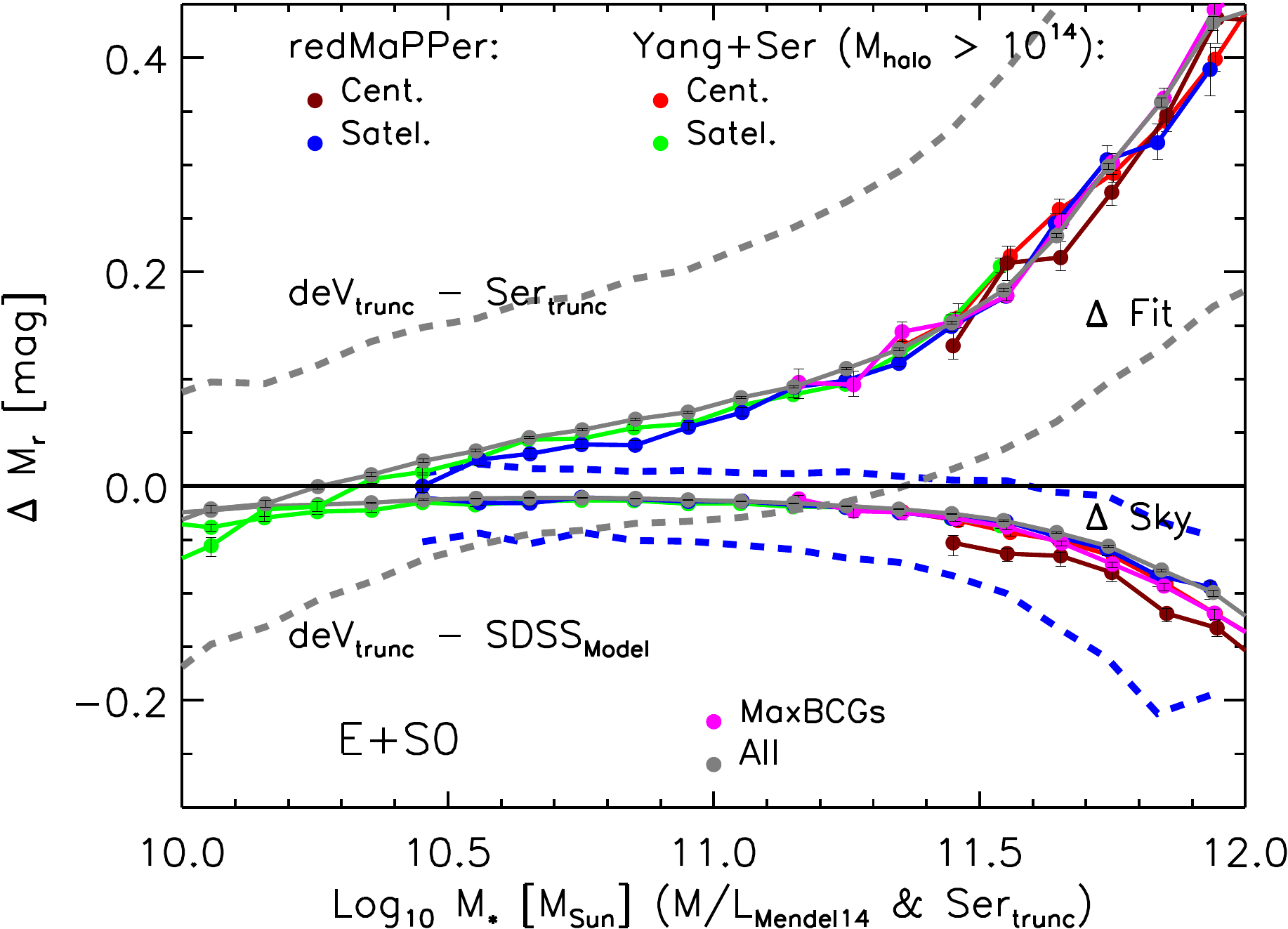} 
 \caption{Same as Figure~\ref{LdiffSer} but now versus $M_*$ rather than absolute magnitude. }
 \label{MdiffSer}
\end{figure*}

The main text studied differences between {\tt PyMorph SerExp} and SDSS pipeline photometry.  Here we show the corresponding results if we use (truncated) {\tt PyMorph Ser} instead of (truncated) {\tt SerExp} values.  Meert et al. (2013) and Bernardi et al. (2014) have argued that the {\tt SerExp} photometry we present in the main text is more unbiased than the {\tt Ser} photometry here.  However, single {\tt Ser} fits are more common in the literature, which is why we show them here.  

Figure~\ref{LdiffSer} shows that the differences with respect to SDSS sky are similar to when fitting {\tt SerExp} profiles (compare Figure~\ref{Ldiffb}), but the remaining difference is substantially larger than before.  In addition, even though the differences are larger, centrals and satellites are similar when shown as a function of {\tt PyMorph} photometry but different when shown as a function of {\tt Model} magnitudes (left and right panels, respectively), as was the case in Figure~\ref{Ldiffb}.  Figure~\ref{MdiffSer} shows that this is also true if we use {\tt Ser} photometry to determine the stellar masses.

\section{Selection effects for analyses at fixed halo mass}\label{secbias}
\begin{figure}
 \centering
 \includegraphics[scale = .42]{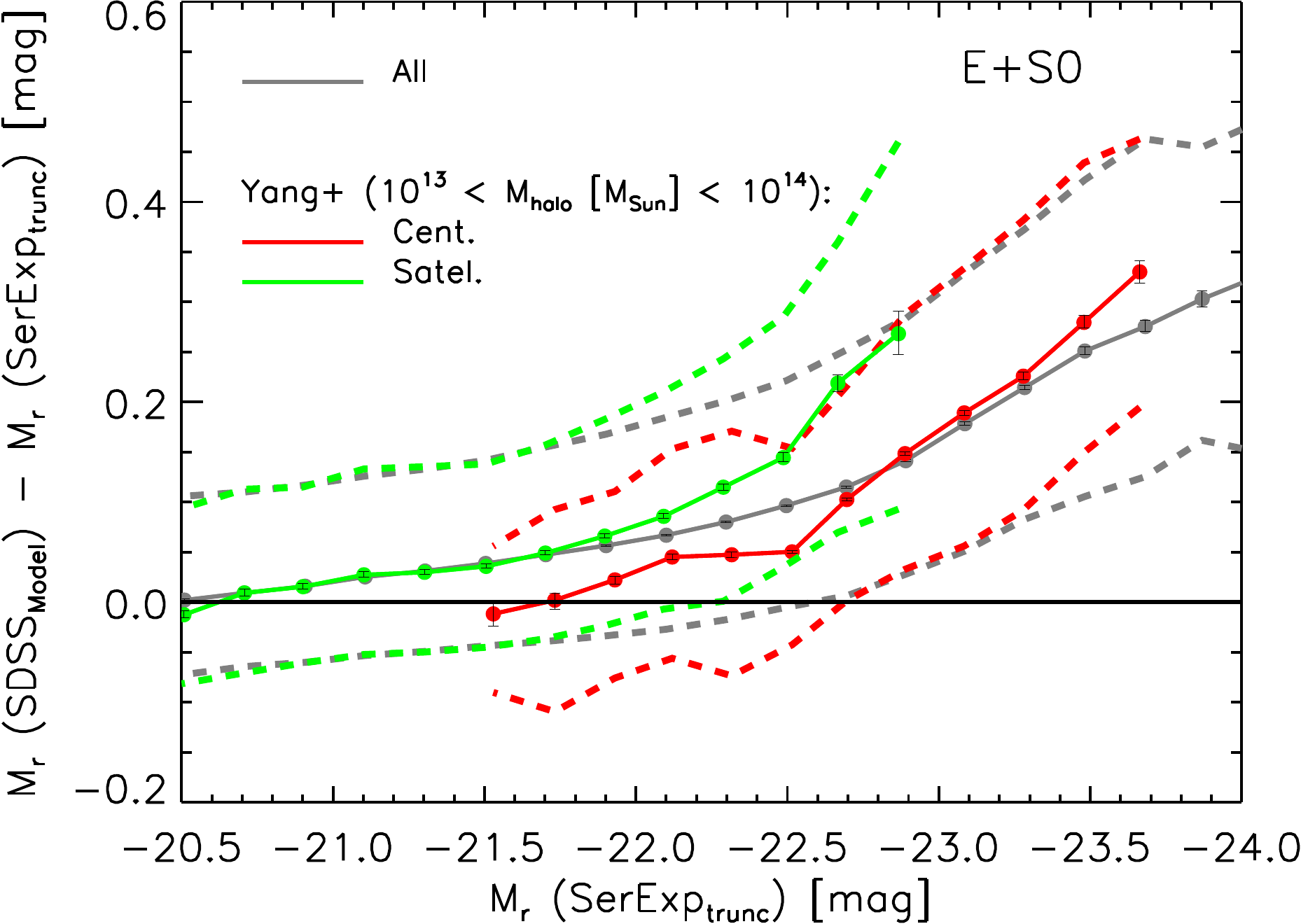}
 \caption{Same as Figure~\ref{lowMhalo}, except that the difference between {\tt Model} and {\tt SerExp} (truncated) magnitudes for galaxies in {\tt Yang+} having group masses between $10^{13}M_\odot$ and $10^{14}M_\odot$, is shown as a function of {\tt SerExp} magnitudes.  The median difference defined by all the E+S0 galaxies (grey; same as corresponding curve in Figure~\ref{lowMhalo}) is significantly different from zero. The curvature away from this median relation, for the brightest satellites (green) and faintest centrals (red), is a selection effect arising from the fact that $M_{\rm Halo}$ is strongly correlated with {\tt Model} magnitudes, but there is scatter in the {\tt Model-SerExp} relation. }
 \label{lowMhaloBias}
\end{figure}

We noted in the main text that it was important to present results using the same photometry that was used to define the centrals, satellites, and halo masses in the {\tt Yang+} group catalog.  Figure~\ref{lowMhalo} shows the result of plotting the PyMorph-SDSS difference versus {\tt Model} magnitude.  Figure~\ref{lowMhaloBias} shows that the scalings change dramatically when we plot versus {\tt SerExp} instead.  The faintest satellites (green) and brightest centrals (red) are remarkably similar to the average (grey), but the brightest satellites and faintest centrals curve away.  We see similar effects if we replace the two-component {\tt SerExp} with single-component {\tt Ser} photometry, and if we use $M_* = (M_*/L)\,L_{\rm Ser}$ or $(M_*/L)\,L_{\rm SerExp}$ instead of $(M_*/L)\,L_{\rm Model}$.  We now show that these offsets are selection effects which result from the fact that the group catalog was constructed using {\tt Model} magnitudes.  For brevity, we only consider the trends for centrals.

Let $m_i$ denote the {\tt model} magnitude of galaxy $i$, $s_i$ its {\tt SerExp} magnitude, and $h_i$ the mass of the halo to which it belongs.  Let $n(m,H)$ denote the number density of objects which have {\tt Model} magnitude $m$ and reside in halos with mass $H\equiv (h_{\rm min}\le h\le h_{\rm max})$.  Then, 
\begin{equation}
 n(m,H) = n(m) \int_H {\rm d}h\, p(h|m) ,
\end{equation}
where the $H$ indicates the limited range in halo masses.  The measurement in Figure~\ref{lowMhalo} is 
\begin{eqnarray}
 \bigl\langle s-m|m,H\bigr\rangle 
 &=& \frac{n(m)}{n(m,H)}\int_H {\rm d}h\, p(h|m)\nonumber\\
  && \quad \times \ \int {\rm d}s\, p(s|m,h)\, (s-m).
\end{eqnarray}
If the scatter between $s$ and $m$ does not depend on halo mass, then we can set $p(s|m,h) = p(s|m)$, and so 
\begin{equation}
 \bigl\langle s-m|m,H\bigr\rangle = \int {\rm d}s\, p(s|m)\, (s-m).
 \label{nobias}
\end{equation}
This is the same for all choices of $H$, and is consistent with the fact that all the curves in Figure~\ref{lowMhalo} are the same.

In contrast, the measurement in Figure~\ref{lowMhaloBias} is 
\begin{align}
 \bigl\langle s-m|s,H\bigr\rangle 
  &= \int_H {\rm d}h \int {\rm d}m\,\frac{n(m,s,h)}{n(s,H)}\ (s-m)\\
  &= \frac{n(s)}{n(s,H)} \int_H {\rm d}h\, p(h|s) \int {\rm d}m\,p(m|s,h)\,(s-m) \nonumber
\end{align}
where 
\begin{equation}
 n(s,H) = \int_H \!\! {\rm d}h \int {\rm d}m \, n(m,s,h) 
        = n(s) \int_H\!\! {\rm d}h\, p(h|s).
\end{equation}
To gain insight, suppose that $p(m|s,h)$ is Gaussian.  Then the mean 
 $\langle m|s,h\rangle$ can always be written as $\langle m|s\rangle + C\,[h-\langle h|s\rangle]$, 
where $C$ does not depend on either $s$ or $h$.  Then 
\begin{equation}
 \bigl\langle s-m|s,H\bigr\rangle 
 = s - \langle m|s\rangle - \frac{n(s)\,C}{n(s,H)} \int_H {\rm d}h\, p(h|s)\,[h-\langle h|s\rangle]. 
 \label{bias}
\end{equation}
The second term is zero if $C=0$; this only happens if the $mh$ correlation is entirely due to the $ms$ and $sm$ correlations.  If $C\ne 0$, then the second term equals zero only if $H$ allows the full range of halo masses.  However, if $H$ selects only a subset of halos, then the integral is non-vanishing, yielding a correction factor which depends on $H$.  If $p(s|h)$ were also Gaussian, then the integral above would equal $\sigma_{h|s}\,[\exp(-y_{\rm min}^2/2)-\exp(-y_{\rm max}^2/2)]/\sqrt{2\pi}$ and the denominator would be $[{\rm erf}(y_{\rm max}/\sqrt{2})-{\rm erf}(y_{\rm min}/\sqrt{2})]/2$, where $y_{\rm max} = (h_{\rm max}-\langle h|s\rangle)/\sigma_{h|s}$ and similarly for $y_{\rm min}$.

\begin{figure}
 \centering
 \includegraphics[scale = .4]{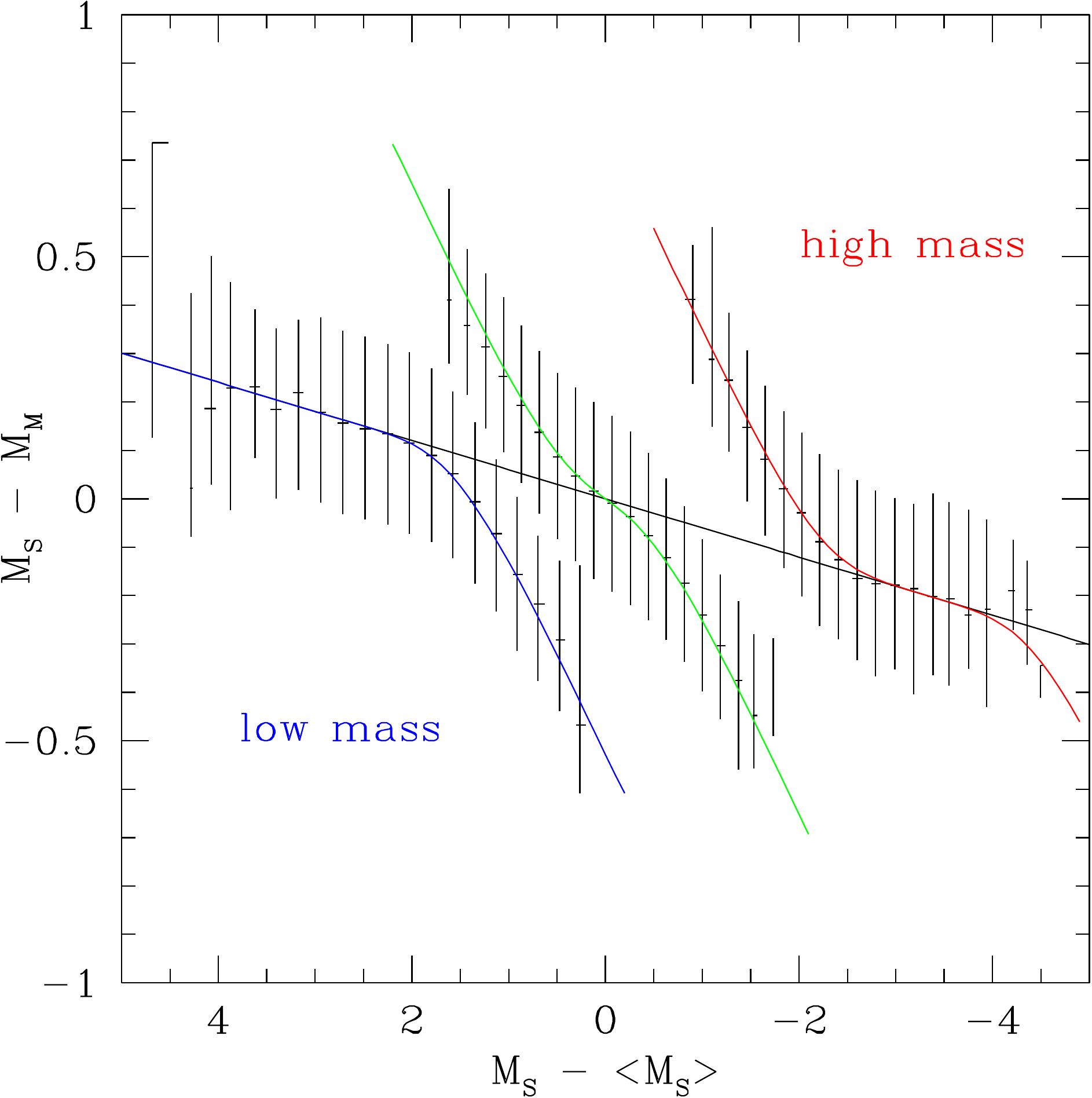}
 \caption{Monte-Carlo demonstration of the selection effect associated with plotting versus the variable on which the group catalog was not defined.  The offsets from the straight line, which are qualitatively similar to those in Figures~\ref{lowMhaloBias} and~\ref{lowMhalo2Bias}, are selection effects which our simple model, equation~(\ref{bias}) describes well.  The offsets depend on halo mass as well as on the range of halo masses included:  the bin width is widest for the lowest mass bin and narrowest for the bin in the middle.  }
 \label{mc}
\end{figure}

To test this, we assumed that the distribution of $m$ is Gaussian with unit variance, that $p(s|m)$ is Gaussian with mean $1.025m$ and rms 0.2, and $p(h|m)$ is Gaussian with mean $-0.4m$ and rms 0.1.  While these scalings are not exactly the same as those in the data, they are similar.  Note in particular that $h$ is directly related to $m$; it only correlates with $s$ because $s$ too is directly related to $m$.  (I.e., the factor $C$ is proportional to the $hm$ correlation.)

We first made Monte-Carlo realizations of this joint distribution of $m$, $s$, and $h$, and checked that measurements of $\bigl\langle s-m|m,H\bigr\rangle$ do not depend on $H$, in good agreement with equation~(\ref{nobias}).
Figure~\ref{mc} shows a similar analysis of $\bigl\langle s-m|s,H\bigr\rangle$; i.e., when we plot the difference between $s$ and $m$ versus $s$ rather than $m$.  The straight black line shows $s - \langle m|s\rangle$ as a function of $s$, and the other three curves show equation~(\ref{bias}) with $(h_{\rm min},h_{\rm max}) = (-5,-0.5)$, $(-0.25,0.25)$, and $(0.75,1.75)$.  We have chosen the different bin centers to illustrate the mass dependence of the offsets, and the different bin widths to show that these offsets appear primarily at the edges of the bins.  The error bars show the corresponding measurements of $\bigl\langle s-m|s,H\bigr\rangle$ in the Monte-Carlo realizations in which $\bigl\langle s-m|m,H\bigr\rangle$ showed no offsets.  The agreement between the measurements and equation~(\ref{bias}) indicates that we understand the origin of the offsets from the straight line.

\begin{figure}
  \centering
  \includegraphics[scale = .42]{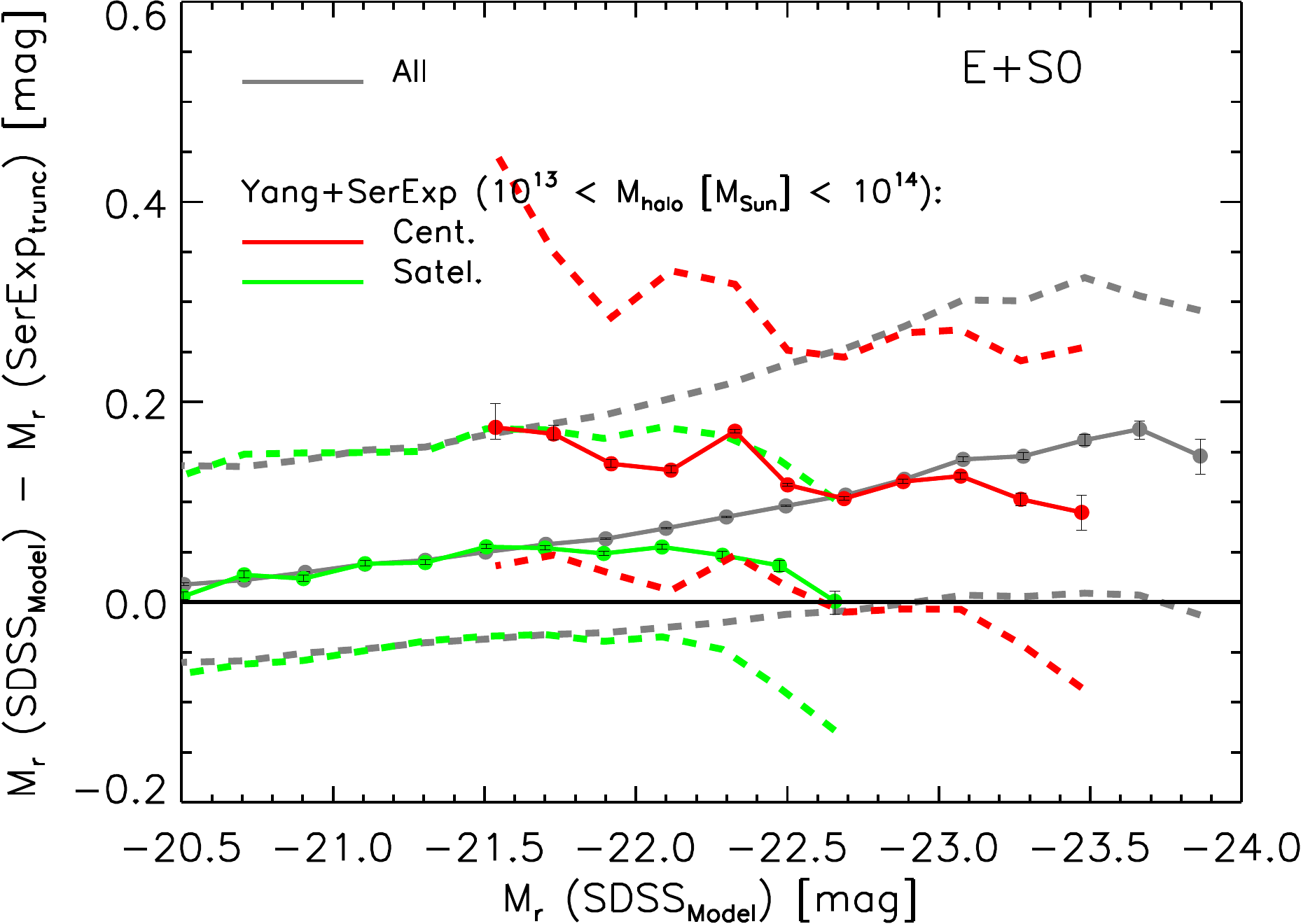}
  \caption{Same as Figure~\ref{lowMhalo2}, but now showing the magnitude difference as a function of {\tt Model} magnitude in the {\tt Yang+SerExp} catalog, in which $M_{\rm Halo}$ is strongly correlated with {\tt SerExp}.  As for Figure~\ref{lowMhaloBias}, the curvature away from the median relation shown by the grey curve is a selection effect.}
 \label{lowMhalo2Bias}
\end{figure}

In particular, the fact that we see no offsets when we plot versus $m$ indicates that the offsets here, and in Figure~\ref{lowMhaloBias}, are selection effects -- they are not physical.  Additional evidence that they are selection effects comes from comparison of Figures~\ref{lowMhalo2} and~\ref{lowMhalo2Bias}.  Both use the {\tt Yang+SerExp} catalog defined in Section~\ref{yang+pymorph} for which halo masses were assigned to the {\tt Yang+} groups on the basis of {\tt SerExp} rather than {\tt Model} magnitudes.  In this case, plots as a function of {\tt SerExp} magnitudes were well-behaved (Figure~\ref{lowMhalo2}), whereas those for {\tt Model} show offsets (Figure~\ref{lowMhalo2Bias}).  I.e., the selection effect is reversed, as expected based on the analysis above.

\end{document}